\documentclass{article}

\usepackage{amssymb}

\newtheorem{theorem}{Theorem}
\newtheorem{lemma}[theorem]{Lemma}
\newtheorem{example}[theorem]{Example}
\newtheorem{definition}[theorem]{Definition}
\newtheorem{corollary}[theorem]{Corollary}
\newtheorem{remark}[theorem]{Remark}
\newtheorem{claim}[theorem]{Claim}

\newenvironment{program}{\tt \begin{tabbing}1234\= 123 \= {\tt
      12}\= 1 \kill}{\end{tabbing}}

\newcommand{\la}{\ensuremath{\:\leftarrow\:}}

\newcommand{\U}{{\ensuremath{\:\cup\:}}}

\newcommand{\ol}[1]{{{\bf #1}}}

\newenvironment{proof}{\noindent {\bf Proof.}}%
{\hfill{$\square$}}    

\title{Properties of Input-Consuming Derivations}

\author{Annalisa Bossi$^1$, Sandro Etalle$^2$, Sabina Rossi$^1$\\[5mm]
$^1$Dipartimento di Informatica,
  Universit\`a di  Venezia \\
  via Torino 155,
  30172 Venezia, Italy
\\[5mm]
$^2$Department of Computer Science,
  University of Maastricht\\
  P.O. Box 616,
  6200 MD Maastricht, The Netherlands\\
  and\\
  CWI -- Center for Mathematics and Computer Science,\\
  P.O.\ Box 94079, 1090 GB Amsterdam, The Netherlands}

\begin{document}
\maketitle

\begin{abstract}
  
  We study the properties of input-consuming derivations of moded
  logic programs. Input-consuming derivations can be used to model the
  behavior of logic programs using dynamic scheduling and employing
  constructs such as \emph{delay declarations}.
  
  We consider the class of \emph{nicely-moded} programs and queries.
  We show that for these programs a weak version of
  the well-known \emph{switching
  lemma} holds also for input-consuming derivations.  Furthermore, we
  show that, under suitable conditions,  there
  exists an algebraic characterization of termination
  of input-consuming derivations.

\end{abstract}

%%%%%%%%%%%%%%%%%%%%%%%%
\section{Introduction}
%%%%%%%%%%%%%%%%%%%%%%%%

Most of the recent logic programming languages provide the possibility
of employing \emph{dynamic scheduling}, i.e., a runtime mechanism
determining which atoms in a query are \emph{selectable} and which
ones are not. In fact, dynamic scheduling has  proven to be
useful in a number of applications; among other things, it allows one
to model coroutining, as shown in \cite{Nai92,HL94}, and parallel
executions, as shown in \cite{Nai88}.  

Let us use the following simple examples to show how dynamic scheduling
can be enforced by using \emph{delay declarations} and how it can
prevent nontermination and unnecessary computations.
 Consider the  program
\texttt{APPEND}
\begin{program}
   \> app([ ],Ys,Ys).\\
  \> app([H|Xs],Ys,[H|Zs]) \la app(Xs,Ys,Zs).
\end{program}
together with the  query
\begin{program}
\>  $Q_1:=$  \>  \> app(Xs,[5,6],Ys), app([1,2],[3,4],Xs).
\end{program}
In this query, if we select and resolve the leftmost atom, we could
easily have to face one of the following two problems. First, the
possibility of nontermination: This is the case if we repeatedly
resolve the leftmost atom against the second clause.  The second problem is
that of inefficiency. If, for instance, in $Q_1$ we resolve the
leftmost atom against the first clause, we obtain the query
\texttt{app([1,2],[3,4],[~])}. This will eventually fail, yiel\-ding
to (unnecessary) backtracking. Notice that if one employs the
rightmost selection rule, $Q_1$ would terminate with success and
without backtracking.  Basically, the problem when selecting
\texttt{app(Xs,[5,6],Ys)}, is that we do not know which clause we
should use for resolving it, and the only practical way for getting to
know this is by waiting until the outermost functor of \texttt{Xs} is
known:  If it is the empty list \texttt{[ ]} we know that we should use
the first clause, if it is the list-constructor symbol we know that we
should use the second clause, if it is something else again, we know then
that the query \emph{fails}.
Notice that the same problems arise for the query

\begin{program}
\>$Q_2:=$ \> \>    app([1,2],[3,4],Xs), app(Xs,[5,6],Ys).
\end{program}
if the rightmost selection rule is considered.

This shows the usefulness of a mechanism for preventing the selection
of those atoms which are not sufficiently instantiated.  
Such a mechanism is in fact offered by most modern languages: In GHC
\cite{Ueda88} programs are augmented with \emph{guards} in order to
control the selection of atoms dynamically.  Moded Flat GHC
\cite{UM94} uses an extra condition on the input positions, which is
extremely similar to the concept of input-consuming derivation step we
refer to the sequel: The resolution of an atom with a definition must
not instantiate the input arguments of the resolved atom. On the other
hand, G\"{o}del \cite{HL94} and ECLiPSe \cite{WNS97} use \emph{delay
  declarations}, and SICStus Prolog \cite{Sicstus97} employs {\tt
  block} declarations (which are a special kind of delay
declarations).  Both {\em delay} and {\tt block} declarations check
the partial instantiation of some arguments of calls.  For instance,
the standard delay declaration for \texttt{APPEND} is
\begin{program}
\> $d_1 :=$ \> \>delay app(Ls, \_, \_) until nonvar(Ls).
\end{program}
This declaration forbids the selection of an atom of the form
$\texttt{app}(s,t,u)$ unless $s$ is a non-variable term, which is
precisely what we need in order to run the queries $Q_1$ or $Q_2$
efficiently.
\\

The
adoption of dynamic scheduling has the disadvantage that various
program properties that have been proven for logic and pure Prolog
programs do not
apply any longer.

The goal of our research is the study of termination properties. This
is motivated by the fact that most of the literature on termination of
logic programs (see De Schreye and Decorte \cite{DD94} for a
survey on this subject) assumes the standard Prolog selection rule,
i.e., the leftmost one.  Notable exceptions are Bezem
\cite{Bez93} and Cavedon \cite{Cav89} who provide results
for all selection rules.  There are only few authors who tackled the
specific problem of verifying the termination of logic programs with
dynamic scheduling.  Namely, Apt and Luitjes \cite{AL95},
Marchiori and Teusink \cite{MT95} and Smaus \cite{Sma99-ICLP}.
We compare our results with the ones in \cite{AL95,MT95,Sma99-ICLP} in the
concluding section.

Another feature of logic programs which does not hold in presence of
dynamic scheduling is the well-known \emph{switching lemma}, which is,
for instance, at the base of the result on the independence of the
selection rule. In this paper we show that -- under certain conditions
-- a weak form of the well-known switching lemma holds.
\\

In order to recuperate at least part of the declarative reading of
logic programming, we follow here the same approach to dynamic
scheduling as \cite{Sma99-ICLP} and we substitute the use of delay
declarations by the restriction to \emph{input-consuming}
de\-ri\-va\-tions.  The definition of input-consuming derivation is
done in two phases. First we give the program a \emph{mode}, that is,
we partition the positions of each atom into \emph{input} and
\emph{output} positions. Then, in presence of modes,
\emph{input-consuming} derivation steps are precisely those in which
the input arguments of the selected atom will not be instantiated by
the unification with the clause's head.  If in a query no atom is
resolvable via an input-consuming derivation step and a failure does
not arise then we have a \emph{deadlock} situation\footnote{As we
  discuss in Section~\ref{sec:ic-dd}, 
this notion of deadlock differs, in some way, from
  the usual one, which is given in the case of programs employing
  delay declarations.}.

For example, the standard mode for the program \texttt{APPEND}
reported above, when used for concatenating two lists, is
\texttt{app(In,In,Out)}.  Notice that in this case the delay
declaration  $d_1$ serves precisely the purpose of guaranteeing
that if an atom of the form $\mathtt{app}(s,t,X)$ (with $X$ being a
variable) is selectable and unifiable with a clause head, then the
resulting derivation step is input-consuming.

It is also worth remarking that, as a large body of literature shows,
the vast majority of ``usual'' programs are actually moded and are,
in a well-defined sense consistent wrt.\ to their modes (e.g., 
well-moded, nicely-moded, simply-moded, etc.); see for
example \cite{AP94a,AM94}, or more simply, the tables of programs we
report in Section~\ref{sec:applicability}, or consider for instance
the logic programming language Mercury \cite{SHC96}, which requires
that its programs are moded (and well-moded).

\subsection*{Contributions of this paper}

In this paper we study some properties of input-consuming derivations.

In the first place we show that, if we restrict ourselves to programs
and queries which are nicely-moded, then  
a weak form of the well-known switching lemma holds.

Furthermore, we study the termination properties of input-consuming 
derivations.  For this we
define the class of \emph{input terminating} programs which
characteri\-zes programs whose input-consuming derivations starting in
a nicely-moded query are finite. In order to prove that a program is
input terminating, we use the concept of \emph{quasi recurrent} program
(similar to, but noticeably less restrictive than the concept of
semi-recurrent program introduced in
\cite{AP94}).  We show that if $P$ is nicely-moded and quasi recurrent
then all its input-consuming derivations starting from a nicely-moded
query terminate.

Furthermore, we demonstrate that under mild additional constraints
(namely, simply-modedness and input-recurrency) the above condition is
both sufficient and \emph{necessary} for ensuring that all
input-consuming derivations starting from a nicely-moded query
terminate.

This approach generalizes the method described in
\cite{Sma99-ICLP} in two ways: First because we also provide conditions
which are both necessary and sufficient, and secondly because we do
not require programs and queries to be well-moded; we only assume that
they are nicely-moded. This is actually crucial: When programs and
queries are well-moded, derivations cannot \emph{deadlock}. Thus, as
opposed to \cite{Sma99-ICLP}, our results capture also termination by
deadlock. For instance, we can easily prove that the query
$\mathtt{app}(X,Y,Z)$ terminates. A more detailed comparison is
presented in the concluding section.

We also show that the results presented in this paper can be extended
to programs and queries which are \emph{permutation} nicely-
or simply-moded,  \cite{SHK98}.

To evaluate the practicality of the results we present, we consider 
the programs from various well-known collections, and we
check whether they satisfy the conditions of our main theorem.
 
The paper is organized as follows. Section \ref{prel} contains some
preliminary notations and definitions.  In Section \ref{sec:input}
input-consuming derivations are introduced and some properties of them
are proven.  In Section 
\ref{sec:switching} we prove that, for nicely-moded input-consuming
programs, a left switching lemma holds.  In Section
\ref{input-ter} a method for proving input termination of programs is
presented, first in a non-modular way, then for modular programs.
In Section \ref{sec:necessity} we show that  this method is necessary for the class of
simply-moded and input-recursive programs.
Section \ref{sec:applicability}
discusses the applicability of our results 
 through simple examples of programs 
and  reports the results obtained by
applying our method to various benchmarks. 
Finally, Section~\ref{sec:conclusion} concludes the paper.

%%%%%%%%%%%%%%%%%%%%%%%%
\section{Preliminaries}
\label{prel}
%%%%%%%%%%%%%%%%%%%%%%%%

The reader is assumed to be familiar with the terminology  and the basic 
results of  logic programs \cite{Apt90,Apt97,Llo87}.

%%%%%%%%%%%%%%%%%%%%%%%%%%%%%%%%%%%%%%
\subsection{Terms and Substitutions}
%%%%%%%%%%%%%%%%%%%%%%%%%%%%%%%%%%%%%%
\noindent
Let ${\cal T}$ be the set of terms built on a finite set 
of \emph{data constructors} ${\cal C}$ and a denumerable set 
of \emph{variable symbols} ${\cal V}$.
A \emph{substitution} $\theta$ 
is a mapping from ${\cal V}$ to $ {\cal T}$ such that 
$\mathit{ Dom}(\theta)=\{X|\;\theta(X)\not = X\}$ is finite. 
For any syntactic object $o$, we denote by 
$\mathit{ Var}(o)$ the set of variables occurring in $o$.
A syntactic object is linear if every variable occurs in it at
most once.
We denote by $\epsilon$  the empty substitution. 
The \emph{composition}
 $\theta\sigma$ of the substitutions $\theta$ and
$\sigma$ is defined as the functional composition, i.e., $\theta \sigma(X)=
\sigma(\theta(X))$.
We consider the pre-ordering $\leq$ (more general than) on
substitutions such that $\theta \leq \sigma$ iff there exists $\gamma$
such that $\theta \gamma=\sigma$.  The result of the application of a
substitution $\theta$ to a term $t$ is said an \emph{instance} of $t$
and it is denoted by $t\theta$. We also consider the pre-ordering
$\leq$ (more general than) on terms such that $t\leq t'$ iff there
exists $\theta$ such that $t\theta=t'$.  We denote by $\approx$ the
associated equivalence relation ({\it variance}).  A substitution
$\theta$ is a \emph{unifier} of terms $t$ and $t'$ iff
$t\theta=t'\theta$. We denote by $\emph{mgu}(t , t')$ any \emph{most
  general unifier} (\emph{mgu}, in short) of $t$ and $t'$.  An mgu
$\theta$ of terms $t$ and $t'$ is called relevant iff $\mathit{
  Var}(\theta)\subseteq \mathit{ Var}(t)\U\mathit{ Var}(t')$.

%%%%%%%%%%%%%%%%%%%%%%%%%%%%%%%%%%%%%%
\subsection{Programs and Derivations}
%%%%%%%%%%%%%%%%%%%%%%%%%%%%%%%%%%%%%%
\noindent
Let ${\cal P}$ be a finite set of \emph{predicate symbols}.  An
\emph{atom} is an object of the form $p(t_1,\ldots,t_n)$ where $p\in
{\cal P}$ is an $n$-ary predicate symbol and $t_1,\ldots,t_n\in {\cal
  T}$.  Given an atom $A$, we denote by $\mathit{ Rel}(A)$ the
predicate symbol of $A$.  A \emph{query} is a finite, possibly empty,
sequence of atoms $A_1,\ldots ,A_m$.  The empty query is denoted by
$\square$.  Following the convention adopted in \cite{Apt97}, we use bold
characters to denote queries.  A \emph{clause} is a formula
$H\leftarrow \mathbf{ B}$ where $H$ is an atom (the \emph{head}) and
$\mathbf{ B}$ is a query (the \emph{body}).  When $\mathbf{ B}$ is
empty, $H\leftarrow \mathbf{ B}$ is written $H\leftarrow $ and is
called a \emph{unit clause}.  A \emph{program} is a finite set of
clauses.  We denote atoms by $A,B,H,\ldots,$ queries by $Q,\mathbf{
  A},\mathbf{ B}, \mathbf{ C}, \ldots,$ clauses by $c,d,\ldots,$ and
programs by~$P$.

Computations are constructed as sequences of ``basic'' steps.
Consider a non-empty query $\mathbf{ A},B,\mathbf{ C}$ and a clause
$c$. Let $H\leftarrow \mathbf{ B}$ be a variant of $c$ variable
disjoint from $\mathbf{ A},B,\mathbf{ C}$. Let $B$ and $H$ unify with
mgu $\theta$.  The query $(\mathbf{ A},\mathbf{ B},\mathbf{ C})\theta$
is called a \emph{resolvent of} $\mathbf{ A},B,\mathbf{ C}$ \emph{and}
$c$ \emph{with selected atom} $B$ \emph{and mgu} $\theta$.  A
\emph{derivation step} is denoted by
\[\mathbf{ A},B,\mathbf{ C}
\stackrel{\theta}\Longrightarrow_{P,c}(\mathbf{ A},\mathbf{
  B},\mathbf{ C})\theta\]
The clause $H\leftarrow \mathbf{ B}$ is called its
\emph{input clause}.  The atom $B$ is called the \emph{selected atom}
of $\mathbf{ A},B,\mathbf{C}$.

If $P$ is clear from the context or $c$ is irrelevant then we drop the
re\-fe\-rence to them.  A derivation is obtained by iterating
derivation steps. A maximal sequence 
%of derivation steps 
\[\delta:=Q_0\stackrel{\theta_1}\Longrightarrow_{P,c_1}Q_1
\stackrel{\theta_2}\Longrightarrow_{P,c_2} \cdots
Q_n\stackrel{\theta_{n+1}}\Longrightarrow_{P,c_{n+1}}Q_{n+1}\cdots\]
is called a \emph{derivation of $P\U\{Q_0\}$}
provided that for eve\-ry step the standardization apart condition
holds, i.e., the input clause employed is variable disjoint from the
initial query $Q_0$ and from the substitutions and the input clauses
used at earlier steps.

Derivations can be finite or infinite.  If
$\delta:=Q_0\stackrel{\theta_1}\Longrightarrow_{P,c_1}\cdots
\stackrel{\theta_n}\Longrightarrow_{P,c_n} Q_n$ is a finite prefix of
a derivation, also denoted $\delta:=Q_0\stackrel{\theta}\longmapsto
Q_n$ with $\theta=\theta_1 \cdots \theta_n$, we say that $\delta$ is a
\emph{partial derivation} and $\theta$ is a \emph{partial computed
  answer substitution} 
of $P\U\{Q_0\}$.  If $\delta$ is maximal and ends with the empty
query then $\theta$ is called \emph{computed answer substitution}
(\emph{c.a.s.}, for short).  The length of a (partial) derivation
$\delta$, denoted by $\mathit{len}(\delta)$, is the number of
derivation steps in $\delta$.

The following definition 
of $\ol B$-step  is due to Smaus \cite{Sma99a}.

\begin{definition}[\ol B-step]
Let $\mathbf{ A},B,\mathbf{ C}
\stackrel{\theta}\Longrightarrow (\mathbf{ A},\mathbf{
  B},\mathbf{ C})\theta$ be a derivation step. We say that
each atom in $\ol B\theta$ is a \emph{direct descendant of B},
and for each atom $E$ in $(\ol A,\ol C)$, $E\theta$ is
a \emph{direct descendant of E}. We say that $E$ is a descendant of $F$
if the pair $(E,F)$ is in the reflexive, transitive closure of the relation
\emph{is a direct descendant of}.
Consider a derivation 
$Q_0\stackrel{\theta_1}\Longrightarrow\cdots
\stackrel{\theta_i}\Longrightarrow Q_i \cdots
\stackrel{\theta_j}\Longrightarrow Q_j\stackrel{\theta_{j+1}}\Longrightarrow 
Q_{j+1}\cdots$. We say that $ Q_j\stackrel{\theta_{j+1}}\Longrightarrow 
Q_{j+1}\cdots$ is a $\ol B$-step if $\ol B$ is a subquery of $Q_i$ and
the selected atom in $Q_j$ is a descendant of an atom in $\ol B$.
\end{definition}

%%%%%%%%%%%%%%%%%%%%%%%%%
\section{Modes and Input-Consuming Derivations}
\label{sec:input}
%%%%%%%%%%%%%%%%%%%%
In this section we introduce the concept of input-consuming derivation
which is strictly related to the notion of mode; we discuss the relations
between input-consuming derivations and programs using delay declarations;
we recall the notion of nicely-moded program
and state some properties.

\subsection{Input-Consuming Derivations} 
\label{sec:ic}
\noindent
Let us first recall the notion of mode.  A \emph{mode} is a function
that labels as \emph{input} or \emph{output} the positions of each
pre\-di\-cate in order to indicate how the arguments of a predicate
should be~used.

\begin{definition}[Mode]
  Consider an $n$-ary predicate symbol $p$.  A \emph{mode} for $p$
  is a function $m_p$ from $\{1,\ldots,n\}$ to $\{\mathit{ In},\mathit{
    Out}\}$.
\end{definition}

\noindent
If $m_p(i)=\mathit{ In}$ (resp.\ \emph{Out}), we say that $i$ is an $\mathit{
  input}$ (resp.\ \emph{output}) \emph{position of} $p$ (wrt.\
 $m_p$). We assume that each predicate symbol has a unique mode
associated to it; multiple modes may be obtained by simply renaming
the predicates.

If $Q$ is a query, we denote by $\mathit{ In}(Q)$ (resp.\ $\mathit{ Out}(Q)$)
the sequence of terms filling in the input (resp.\ output) positions of
predicates in $Q$.  Moreover, when writing an atom as $p(\mathbf{ s},\mathbf{
  t})$, we are indicating with $\mathbf{ s}$ the sequence of terms filling
in the input positions of $p$ and with $\mathbf{ t}$ the sequence of terms
filling in the output positions of $p$.

The notion of input-consuming derivation was introduced in
\cite{Sma99-ICLP} and is defined as follows.

\begin{definition}[Input-Consuming]
\mbox{}
\begin{itemize}
\item[$\bullet$] An atom $p(\mathbf{ s},\mathbf{ t})$ is called
  \emph{input-consuming resolvable wrt.\ a clause $c:=p(\mathbf{
      u},\mathbf{ v})\leftarrow Q$ and a substitution $\theta$} iff
  $\theta=\mathit{ mgu}( p(\mathbf{ s},\mathbf{ t}) , p(\mathbf{
    u},\mathbf{ v}))$ and $\mathbf{ s}=\mathbf{ s}\theta$.
\item[$\bullet$] A derivation step
$$\mathbf{ A},B,\mathbf{ C} \stackrel{\theta}\Longrightarrow_{c}
(\mathbf{ A},\mathbf{ B},\mathbf{ C})\theta$$
 is called
\emph{input-consuming} iff 
the selected atom $B$ is  input-consuming
resolvable wrt.\ the input clause $c$ and the substitution $\theta$.
\item[$\bullet$] A derivation  is called
\emph{input-consuming} iff all its
 derivation steps are
 input-consuming.
\end{itemize}
\end{definition}
The following lemma states that we are allowed to restrict our
attention to input-consuming derivations with relevant mgu's.
\begin{lemma}
\label{relevance}
Let $p(\mathbf{ s},\mathbf{ t})$ and $p(\mathbf{ u},\mathbf{ v})$ be
two atoms.  If there exists an mgu $\theta$ of $p(\mathbf{ s},\mathbf{
  t})$ and $p(\mathbf{ u},\mathbf{ v})$ such that $\mathbf{
  s}\theta=\mathbf{ s}$, then there exists a \emph{relevant} mgu
$\vartheta$ of $p(\mathbf{ s},\mathbf{ t})$ and $p(\mathbf{
  u},\mathbf{ v})$ such that $\mathbf{ s}\vartheta=\mathbf{ s}$.
\end{lemma}
\begin{proof}
  Since $p(\mathbf{ s},\mathbf{ t})$ and $p(\mathbf{ u},\mathbf{ v})$
  are unifiable, there exists a relevant mgu $\theta_{\mathit{rel}}$
  of them (cfr.  \cite{Apt97}, Theorem 2.16). Now,
  $\theta_{\mathit{rel}}$ is a renaming of $\theta$. Thus
  $\mathbf{s}\theta_{\mathit{rel}}$ is a variant of $\mathbf{ s}$.
  Then there exists a renaming $\rho$ such that $\mathit{
    Dom}(\rho)\subseteq \mathit{ Var}(\mathbf{ s},\mathbf{ t},\mathbf{
    u},\mathbf{ v})$ and $\mathbf{
    s}\theta_{\mathit{rel}}\rho=\mathbf{s}$. Now, take
  $\vartheta=\theta_{\mathit{rel}}\rho$.
\end{proof}
\vspace{3mm}

From now on, we assume that all mgu's used in the input-consuming
derivation steps are relevant.

\begin{example}
\label{i-c}
Consider the program {\tt REVERSE} with accumulator in the modes defined below.
\begin{program}
\> mode reverse(In, Out).\\
\> mode reverse\_acc(In,Out,In)\\[2mm]
\> reverse(Xs,Ys) \la reverse\_acc(Xs,Ys,[ ]).\\
\> reverse\_acc([ ],Ys,Ys).\\
\> reverse\_acc([X|Xs],Ys,Zs)  \la reverse\_acc(Xs,Ys,[X|Zs]).
\end{program}

The derivation  $\delta$ of $\mathtt{ REVERSE}\U\{\mathtt{
  reverse([X1,X2],Zs)}\}$ depicted below is input-consuming.

\begin{program}
\> $\delta :=$  \>  \>  $\mathtt{ reverse([X1,X2],Zs)}\Rightarrow
  \mathtt{ reverse\_acc([X1,X2],Zs,[\;])}\Rightarrow$\\
 \>  \>  \> $\mathtt{ reverse\_acc([X2],Zs,[X1])}\Rightarrow \mathtt{
    reverse\_acc([\;],Zs,[X2,X1])}\Rightarrow \square$.
\end{program}
\end{example}

%%%%%%%%%%%%%%%%%%%%%%%%%%
\subsection{Input-Consuming vs.\ Delay Declarations}
\label{sec:ic-dd}
%%%%%%%%%%%%%%%%%%%%%%%%%%%%%

Delay declarations are by far the most popular mechanism for
implementing dynamic scheduling. However, being a non-logical
mechanism, they are difficult to model and there are few 
proposals concerning their semantics \cite{Mar97} and \cite{FGMP97}.

An alternative approach to dynamic scheduling, which is much more declarative
in nature, has been proposed by Smaus \cite{Sma99-ICLP}. It consists
in the use of input-consuming derivations.

There is a main difference between the concept of delay declaration
and the one of input-consuming derivation: While in the first case
only the atom selectability is controlled, in the second one both
the atom and the clause selectability are affected.
  In fact, in presence of delay
declarations, if an atom is selectable then it can be resolved with
respect to any
program clause (provided it unifies with its head); on the contrary,
in an input-consuming derivation, if an atom is selectable then it is
input-consuming resolvable wrt.\ 
 some, but not necessarily all, program
clauses, i.e, only a restricted class of clauses can be used for
resolution.

 Also the concept of
\emph{deadlock} has to be understood in two different ways. For
programs using delay declarations a deadlock situation occurs when no
atom in a query satisfies the delay declarations (i.e., no atom is
selectable), while for input-consuming derivations a deadlock occurs
when no atom in a query is resolvable via an input-consuming
derivation step and the derivation does not fail, i.e., there is some
atom in the query which unifies with a clause head but the unification
is not input-consuming.

In spite of these differences, in many situations there is a strict
relation between programs using delay declarations and input-consuming
derivations.  This relation is studied by Smaus in his PhD thesis
\cite{Sma99a}. More precisely, Smaus proves a result that relates
\texttt{block} declarations and input-consuming derivations.  A
\texttt{block} declaration is a special case of delay declaration and
it is used to declare that certain arguments of an atom must be
\emph{non-variable} when the atom is selected for resolution.  In
Chapter 7 of \cite{Sma99a}, Smaus shows that \texttt{block}
declarations can be used to ensure that derivations are
input-consuming.  
In force of this result and of practical experience, 
we might  claim that in most ``usual'' moded programs
using them, delay declarations are employed precisely for ensuring the input-consumedness
of the derivations.

 In fact, delay declarations are generally
employed to guarantee that the interpreter will not use an
``inappropriate'' clause for resolving an atom (the other, perhaps
less prominent, use of delay declarations is to ensure absence of
runtime errors, but we do not address this issue in this paper). This
is achieved by preventing the selection of an atom until a certain
degree of instantiation is reached. This degree of instantiation
ensures then that the atom is unifiable only with the heads of the
``appropriate'' clauses. In presence of modes, we can reasonably
assume that this degree of instantiation is the one of the
\emph{input} positions, which are the ones carrying the information.
Now, it is easy to see that a derivation step involving a clause $c$
is input-consuming iff no further instantiation of the input positions
of the resolved atom could prevent it from being resolvable with $c$.
Therefore $c$ must belong to the set of ``appropriate'' clauses for
resolving it. Thus, the concepts of input-consuming derivation and of
delay declarations are often employed for ensuring the same
properties.

\subsection{Nicely-Moded Programs}
In the sequel of the paper we will restrict ourselves to programs and
queries which are nicely-moded.  In this section we report the
definition of this concept together with some basic important
properties of nicely-moded programs.

\begin{definition}[Nicely-Moded]
\mbox{}
\begin{itemize}
\item
A query $Q:=p_1(\mathbf{ s}_1,\mathbf{ t}_1),\ldots,p_n(\mathbf{ s}_n,\mathbf{ t}_n)$
is \emph{nicely-moded} if $\mathbf{ t}_1,\ldots,\mathbf{ t}_n$ is a linear
sequence of terms and for all $i\in\{1,\ldots,n\}$
$$\mathit{ Var}(\mathbf{ s}_i)\cap \bigcup_{j=i}^n\mathit{ Var}(\mathbf{ t}_j)=\emptyset.$$
\item
A clause $c=p(\mathbf{ s}_0,\mathbf{ t}_{0})\leftarrow Q$
is \emph{nicely-moded} if $Q$ is nicely-moded and
$$\mathit{ Var}(\mathbf{ s}_0)\cap \bigcup_{j=1}^n\mathit{ Var}(\mathbf{ t}_j)=\emptyset.$$
In particular, every unit clause is nicely-moded.
\item
A program $P$ is nicely-moded if all of its clauses are nicely-moded.
\end{itemize}
\end{definition}

Note that a one-atom query $p(\mathbf{ s},\mathbf{ t})$ is nicely-moded if and
only if $\mathbf{ t}$ is linear and $\mathit{ Var}(\mathbf{ s})\cap \mathit{ Var}(\mathbf{
  t})=\emptyset$.

\begin{example}
\label{programs}
\mbox{}
\begin{itemize}
\item[$\bullet$]
The program {\tt APPEND} 
in the modes \texttt{app(In,In,Out)} is nicely-moded.

\item[$\bullet$]
The program {\tt REVERSE} 
with accumulator in the modes depicted in  Example~\ref{i-c} is nicely-moded.

\item[$\bullet$] The following program {\tt MERGE}   is nicely-moded.
\begin{program}
\> mode merge(In,In,Out).\\[2mm]
\> merge(Xs,[ ],Xs).\\
\> merge([ ],Xs,Xs).\\
\> merge([X|Xs],[Y|Ys],[Y|Zs])  \la Y < X, merge([X|Xs],Ys,Zs).\\
\> merge([X|Xs],[Y|Ys],[X|Zs])  \la Y > X, merge(Xs,[Y|Ys],Zs).\\
\> merge([X|Xs],[X|Ys],[X|Zs])  \la merge(Xs,[X|Ys],Zs).
\end{program}
\end{itemize}
\end{example} 

The following result is due to Smaus \cite{Sma99a}, and states that the
class of programs and queries we are considering is persistent under
resolution.

\begin{lemma}
\label{Smaus1}
Every resolvent of a nicely-moded query $Q$ and a nicely-moded clause
$c$, where the derivation step is input-consuming and
$\mathit{ Var}(Q)\cap \mathit{ Var}(c)=\emptyset$, is nicely-moded.
\end{lemma}

The following Remark, also in \cite{Sma99a}, is an immediate
consequence of the definition of input-consuming derivation step and
the fact that the mgu's we consider are relevant.

\begin{remark}
\label{Smaus2}
Let the program $P$ and the query $Q:=\mathbf{ A},p(\mathbf{ s},\mathbf{ t}),\mathbf{
  C}$ be nicely-moded. \\
If $\mathbf{ A},p(\mathbf{ s},\mathbf{ t}),\mathbf{
  C}\stackrel{\theta} 
\Longrightarrow (\mathbf{ A},\mathbf{ B},\mathbf{ C})\theta$ is an
input-consuming derivation step with selected atom $p(\mathbf{ s},\mathbf{
  t})$, then $\mathbf{ A}\theta=\mathbf{ A}$.
\end{remark}

%%%%%%%%%%%%%%%%%%%%%%%%%%%%%%
\section{The Left Switching Lemma}
\label{sec:switching}
%%%%%%%%%%%%%%%%%%%%%%%%%%
The \emph{switching lemma} (see for instance \cite{Apt97}, Lemma 3.32) is a
well-known result
which allows one to prove  the independence of the computed answer
substitutions from the selection rule.

In the case of logic programs using dynamic scheduling, the switching
lemma does not hold any longer. For example, in program
\texttt{APPEND} reported in the introduction (together with the delay
declaration $d_1$) we have that the rightmost atom of
$Q_2$ is selectable only after the leftmost one has been
resolved; i.e., the switching lemma cannot be applied.

Nevertheless we can show that, for input-consuming derivations of
nicely-moded programs, a weak version
of the switching lemma still holds.
Intuitively, we show that we  can switch the selection of two
atoms whenever this results in a 
left to right selection.
 For this reason, we call it
\emph{left switching lemma}.

First, we need one technical result, stating that the only variables
of a query that can be ``affected'' in an input-consuming
 derivation process are
those occurring in some output positions. 
Intuitively, 
this means that 
if the input arguments of a call are not ``sufficiently instantiated''
then it is delayed until it allows for an input-consuming
derivation step (if it is not the case then a deadlock situation will arise).

\begin{lemma}
\label{sandro1}
Let the program $P$ and the query $Q$ be nicely-moded.
Let $\delta:=Q\stackrel{\theta}
\longmapsto Q'$ be a partial input-consuming derivation
of $P\U\{Q\}$.
Then, for all $x\in \mathit{ Var}(Q)$ and
$x\not \in\mathit{ Var}(\mathit{ Out}(Q))$, $x\theta=x$.
\end{lemma}
\begin{proof}
Let us first establish the following claim.
\begin{claim}
\label{lin}
Let $\mathbf{ z}$ and $\mathbf{ w}$ be two variable disjoint sequences
of terms such that 
$\mathbf{ w}$ is linear and $\theta=\mathit{ mgu}(\mathbf{ z} , \mathbf{ w})$.
If $s_1$ and $s_2$ are two variable disjoint terms occurring
in $\mathbf{ z}$ then $s_1\theta$ and $s_2\theta$ are variable
disjoint terms.
\end{claim}
\begin{proof}
The result follows from Lemmata 11.4 and 11.5 in \cite{AP94}.
\end{proof}

We proceed with the proof of the lemma by induction on $\mathit{
  len}(\delta)$.

\emph{Base Case}. Let $\mathit{ len}(\delta)=0$.  In this case $Q=Q'$ and
the result follows trivially.

\emph{Induction step}.  Let $\mathit{ len}(\delta)>0$. Suppose that
$Q:=\mathbf{ A},p(\mathbf{ s},\mathbf{ t}), \mathbf{ C}$ and
\[\delta :=\mathbf{ A},p(\mathbf{ s},\mathbf{ t}),
\mathbf{ C} \stackrel{\theta_1}
\Longrightarrow (\mathbf{ A},\mathbf{ B},\mathbf{ C})\theta_1
\stackrel{\theta_{2}}\longmapsto Q'\]
where $p(\mathbf{ s},\mathbf{ t})$ is the selected atom of $Q$,
$c:=p(\mathbf{ u},\mathbf{ v})\leftarrow \mathbf{ B}$ is the input clause used in the 
first derivation step, $\theta_1$ is a relevant
mgu of $p(\mathbf{ s},\mathbf{ t})$ and
$p(\mathbf{ u},\mathbf{ v})$ 
and $\theta=\theta_1\theta_2$.

Let $x\in \mathit{ Var}(\mathbf{ A},p(\mathbf{ s},\mathbf{
  t}),\mathbf{ C})$ and $x\not \in \mathit{ Var}(\mathit{
  Out}(\mathbf{ A},p(\mathbf{ s},\mathbf{ t}),\mathbf{ C}))$.  We
first~show~that
\begin{equation}
\label{eq:xtheta1=x}
x\theta_1=x
\end{equation} 
We distinguish two cases.

$(a)$ $x\in \mathit{ Var}(\mathbf{ s})$.  In this case, property
$(\ref{eq:xtheta1=x})$ follows from the hypothesis that $\delta$ is
input-consuming.

$(b)$ $x\not \in \mathit{ Var}(\mathbf{ s})$. 
Since  $x\in \mathit{ Var}(\mathbf{ A},p(\mathbf{ s},\mathbf{
  t}),\mathbf{ C})$, by standardization apart,
we have that $x\not\in\mathit{ Var}(p(\mathbf{ u},\mathbf{ v}))$. 
Moreover,
since  $x\not \in \mathit{ Var}(\mathit{
  Out}(\mathbf{ A},p(\mathbf{ s},\mathbf{ t}),\mathbf{ C}))$,
it also holds  that 
$x\not\in\mathit{ Var}(p(\mathbf{ s},\mathbf{ t}))$. 
Then, property $(\ref{eq:xtheta1=x})$ 
follows from   relevance of
$\theta_1$.
\\[2mm]
Now we show that
\begin{equation}
\label{eq:xtheta2=x}
x\theta_2=x
\end{equation} 
Again, we distinguish two cases:

$(c)$ $x\not \in \mathit{ Var}((\mathbf{ A},\mathbf{ B},\mathbf{
  C})\theta_1)$.  In this case, because of the standardization apart
condition, $x$ will never occur in $(\mathbf{ A},\mathbf{ B},\mathbf{
  C})\theta_1 \stackrel{\theta_{2}}\longmapsto Q'$. Hence, $x\not \in
\mathit{ Dom}(\theta_2)$ and $x\theta_2=x$.

$(d)$ $x\in \mathit{ Var}((\mathbf{ A},\mathbf{ B},\mathbf{
  C})\theta_1)$.  In this case, in order to prove $(\ref{eq:xtheta2=x})$ 
we show that
$x\not \in \mathit{ Var}(\mathit{ Out}((\mathbf{ A},\mathbf{
  B},\mathbf{ C})\theta_1))$.
The result then follows by the inductive hypothesis.

From the standardization apart, relevance of $\theta_1$ and
the fact that the first derivation step is input-consuming,
 it follows that $\mathit{ Dom}(\theta_1)\cap
\mathit{ Var}(Q)
\subseteq \mathit{ Var}(\mathbf{ t})$.

From the hypothesis that $Q$ is nicely-moded, $\mathit{ Var}(\mathbf{
  t})\cap \mathit{ Var}(\mathit{ Out}(\mathbf{ A},\mathbf{
  C}))=\emptyset$.  Hence, $\mathit{ Var}(\mathit{ Out}(\mathbf{
  A},\mathbf{ C}))\theta_1=\mathit{ Var} (\mathit{ Out}(\mathbf{
  A},\mathbf{ C}))$. Since $x\not\in \mathit{ Var}(\mathit{
  Out}(\mathbf{ A},\mathbf{ C}))$, this proves that $x\not\in \mathit{
  Var}(\mathit{ Out}((\mathbf{ A},\mathbf{ C})\theta_1))$.

It remains to be proven that $x\not \in \mathit{ Var}(\mathit{
  Out}(\mathbf{ B}\theta_1)$.  
We distinguish two cases.

$(d1)$ $x\not \in \mathit{ Var}(\mathbf{ s})$. Since
 $x\not \in \mathit{ Var}(p(\mathbf{ s},\mathbf{
  t}))$,  the fact that
$x\not \in \mathit{
  Var}(\mathit{ Out}(\mathbf{ B}\theta_1)$ follows immediately by 
standardization apart  condition and  relevance of $\theta_1$.

$(d2)$ $x \in \mathit{ Var}(\mathbf{ s})$.  By known results (see
\cite{Apt97}, Corollary 2.25), there exists two relevant mgu
$\sigma_1$ and $\sigma_2$ such that
\begin{itemize}
\item $\theta_1=\sigma_1\sigma_2$,
\item $\sigma_1=\mathit{ mgu}(\mathbf{ s} , \mathbf{ u})$,
\item $\sigma_2=\mathit{ mgu}(\mathbf{ t}\sigma_1 , \mathbf{ v}\sigma_1)$.
\end{itemize}
From relevance of $\sigma_1$ and the fact that,
 by nicely-modedness of
$Q$, 
$\mathit{ Var}(\mathbf{ s})\cap \mathit{ Var}(\mathbf{ t})=\emptyset$,
we have that $\mathbf{ t}\sigma_1=\mathbf{ t}$, and
by the standardization apart condition $\mathit{ Var}(\mathbf{ t})
\cap \mathit{ Var}(\mathbf{ v}\sigma_1)=\emptyset$.
Now by nicely-modedness of $c$,
$\mathit{ Var}(\mathbf{ u})\cap\mathit{ Var}(\mathit{ Out}(\mathbf{ B}))=\emptyset$.
Since $\sigma_1$ is relevant and by the standardization apart
condition it follows that
\begin{equation}
\label{eq:varusigma1inters}
\mathit{ Var}(\mathbf{ u}\sigma_1)\cap\mathit{ Var}
(\mathit{ Out}(\mathbf{ B}\sigma_1))=\emptyset
\end{equation}
The proof proceeds now by contradiction.  Suppose that $x\in \mathit{
  Var}(\mathit{ Out}(\mathbf{ B}\sigma_1\sigma_2))$.  Since by
hypothesis $x\in \mathit{ Var}(\mathbf{ s})$, and $\mathbf{
  s}=\mathbf{ u}\sigma_1\sigma_2$, we have that $\mathit{
  Var}(\mathbf{ u}\sigma_1\sigma_2)\cap\mathit{ Var} (\mathit{
  Out}(\mathbf{ B}\sigma_1\sigma_2))\not=\emptyset$.  By
$(\ref{eq:varusigma1inters})$, this means that there exist two
distinct variables $z_1$ and $z_2$ in $\mathit{ Var}(\sigma_2)$ such
that $z_1\in \mathit{ Var}(\mathit{ Out}(\mathbf{ B}\sigma_1))$,
$z_2\in \mathit{ Var}(\mathbf{ u}\sigma_1)$ and
\begin{equation}
\label{eq:varz1sigmaz2}
\mathit{ Var}(z_1\sigma_2)\cap
\mathit{ Var}(z_2\sigma_2)\not=\emptyset
\end{equation}
Since, by the standardization apart condition
 and relevance of the mgu's,
 $\mathit{ Var}(\sigma_2) $ $\subseteq \mathit{ Var}(\mathbf{ v}\sigma_1)\U
\mathit{ Var}(\mathbf{ t})$ and
$(\mathit{ Var}(\mathit{ Out}(\mathbf{ B}\sigma_1))
\U \mathit{ Var}(\mathbf{ u}\sigma_1))
\cap \mathit{ Var}(\mathbf{ t})=
\emptyset$,
we have that $z_1$ and $z_2$
are two disjoint subterms of  $\mathbf{ v}\sigma_1$.
Since  $\sigma_2=
\mathit{ mgu}(\mathbf{ t} ,\mathbf{ v}\sigma_1)$,  $\mathbf{ t}$ is linear
and disjoint from $\mathbf{ v}\sigma_1$,
(\ref{eq:varz1sigmaz2}) contradicts  Claim \ref{lin}.
\end{proof}
\vspace{3mm}

The following corollary is an immediate consequence of the above lemma 
and the definition of nicely-moded program.

\begin{corollary}
\label{corollary}
Let the program $P$ and the one-atom query $A$ be nicely-moded.  Let
$\delta:=A\stackrel{\theta} \longmapsto Q'$ be a partial
input-consuming derivation of $P\U\{A\}$.  Then, for all $x\in \mathit{
  Var}(\mathit{ In}(A))$, $x\theta=x$.
\end{corollary}

Next is the main result of this section, showing that for
input-consuming nicely-moded programs one half of the well-known
switching lemma holds.

\begin{lemma}[Left-Switching]
\label{switching}
Let the program $P$ and the query $Q_0$ be nicely-moded. Let
$\delta$ be a partial
 input-consuming derivation of $P\U\{Q_0\}$ of the form
\[\delta :=Q_0\stackrel{\theta_1}
\Longrightarrow_{c_1} Q_1\cdots Q_n
\stackrel{\theta_{n+1}}\Longrightarrow_{c_{n+1}} Q_{n+1} 
\stackrel{\theta_{n+2}}\Longrightarrow_{c_{n+2}} Q_{n+2}\]
where
\begin{itemize}
\item
$Q_n$ is a query of the form $\mathbf{ A},B,\mathbf{ C},D,\mathbf{ E}$,
\item
$Q_{n+1}$ is a resolvent of $Q_n$  and $c_{n+1}$
wrt.\ $D$,
\item
$Q_{n+2}$ is a resolvent of $Q_{n+1}$  and $c_{n+2}$
wrt.\  $B\theta_{n+1}$.
\end{itemize}
Then, there exist $Q'_{n+1}$, $\theta'_{n+1}$, $\theta'_{n+2}$
and a  derivation $\delta'$ such that
\[\theta_{n+1}\theta_{n+2}=\theta'_{n+1}\theta'_{n+2}\]
and
\[\delta' :=Q_0\stackrel{\theta_1}
\Longrightarrow_{c_1} Q_1\cdots Q_n
\stackrel{\theta'_{n+1}}\Longrightarrow_{c_{n+2}} Q'_{n+1} 
\stackrel{\theta'_{n+2}}\Longrightarrow_{c_{n+1}} Q_{n+2}\]
where $\delta'$ is input-consuming and
\begin{itemize}
\item $\delta$ and $\delta'$ coincide up to the resolvent
$Q_n$,
\item 
$Q'_{n+1}$ is a resolvent of $Q_n$  and $c_{n+2}$
wrt.\  $B$,
\item 
$Q_{n+2}$ is a resolvent of $Q'_{n+1}$  and $c_{n+1}$
wrt.\  $D\theta'_{n+1}$,
\item $\delta$ and $\delta'$ coincide after the resolvent
$Q_{n+2}$.
\end{itemize}
\end{lemma}
\begin{proof}
  Let $B:=p(\mathbf{ s},\mathbf{ t})$, $D:=q(\mathbf{ u},\mathbf{ v})$,
  $c_{n+1}:=q(\mathbf{ u'},\mathbf{ v'}) \leftarrow \mathbf{ D}$ and $c_{n+2}:=
  p(\mathbf{ s'},\mathbf{ t'})\leftarrow \mathbf{ B}$.  Hence, $\theta_{n+1} =
  \mathit{ mgu}(q(\mathbf{ u},\mathbf{ v}) , q(\mathbf{ u'},\mathbf{ v'}))$ and
\begin{equation}
\label{eq:u+1=u}
\mathbf{ u}\theta_{n+1}=\mathbf{ u},\mbox{ since $\delta$ is
input-consuming.}
\end{equation}
By $(\ref{eq:u+1=u})$ and the fact that $Q_n$ is nicely-moded and
$\theta_{n+1}$ is relevant, we have that $p(\mathbf{ s},\mathbf{
  t})\theta_{n+1}= p(\mathbf{ s},\mathbf{ t})$.  Then, $\theta_{n+2} =
\mathit{ mgu}(p(\mathbf{ s},\mathbf{ t})\theta_{n+1} , p(\mathbf{
  s'},\mathbf{ t'})) =\mathit{ mgu}(p(\mathbf{ s},\mathbf{ t}) ,
p(\mathbf{ s'},\mathbf{ t'}))$ and
\begin{equation}
\label{eq:s+2=s}
\mathbf{ s}\theta_{n+2}=\mathbf{ s},\mbox{ since $\delta$ is input-consuming.}
\end{equation}
Moreover,\footnote{We use  the notation $\mathit{mgu}(E)$ 
to denote the mgu of a set of equations $E$, see \cite{Apt97}.}
\begin{equation}
\label{eq:thetan+1etc}
\theta_{n+1} \theta_{n+2} = 
\mathit{mgu}\{p(\mathbf{ s},\mathbf{ t})= p(\mathbf{ s'},\mathbf{ t'}),
q(\mathbf{ u},\mathbf{ v})= q(\mathbf{ u'},\mathbf{ v'})\}
 = \theta_{n+2}\theta'_{n+2}
\end{equation}
where 
\[
\theta'_{n+2} 
 =  \mathit{ mgu}(q(\mathbf{ u},\mathbf{ v})\theta_{n+2} , q(\mathbf{ u'},\mathbf{
v'})\theta_{n+2})\\
 =  \mathit{ mgu}(q(\mathbf{ u},\mathbf{ v})\theta_{n+2} ,  q(\mathbf{ u'},\mathbf{ v'}) ).
\]
We construct the derivation $\delta'$ as follows.
\[\delta' :=Q_0\stackrel{\theta_1}
\Longrightarrow_{c_1} Q_1\cdots Q_n
\stackrel{\theta'_{n+1}}\Longrightarrow_{c_{n+2}} Q'_{n+1} 
\stackrel{\theta'_{n+2}}\Longrightarrow_{c_{n+1}} Q_{n+2}
\]
where
\begin{equation}
\label{eq:theta'n+1etc}
\theta'_{n+1}=\theta_{n+2}
\end{equation}
By $(\ref{eq:s+2=s})$, $Q_n
\stackrel{\theta'_{n+1}}\Longrightarrow_{c_{n+2}} Q'_{n+1} $ is an
input-consuming derivation step.  Observe now that
\[ \begin{array}{lllll} \mathbf{ u}\theta'_{n+1}\theta'_{n+2}
  & = & \mathbf{ u}\theta_{n+2}\theta'_{n+2}, & (\mbox{by $(\ref{eq:theta'n+1etc})$})\\
  &=&  \mathbf{ u}\theta_{n+1}\theta_{n+2}, & (\mbox{by $(\ref{eq:thetan+1etc})$})\\
  &=&  \mathbf{ u}\theta_{n+2}, & (\mbox{by $(\ref{eq:u+1=u})$})\\
  &=& \mathbf{ u}\theta'_{n+1}, & (\mbox{by $(\ref{eq:theta'n+1etc})$}).
\end{array}
\]
This proves that 
$Q'_{n+1}
\stackrel{\theta'_{n+2}}\Longrightarrow_{c_{n+1}} Q'_{n+2} $
is an input-consuming derivation step.
\end{proof}
\vspace{3mm}

This result shows that it is always possible to proceed left-to-right
to resolve the selected atoms. Notice that this is different than
saying that the leftmost atom of a query is always resolvable: It can
very well be the case that the leftmost atom is suspended and the one
next to it is resolvable.  However, if the leftmost atom of a query is
not resolvable then we can state that the derivation will not succeed,
i.e., either it ends by deadlock, or by failure or it is infinite.

It is important to notice that if we drop the nicely-modedness
condition the above lemma would not hold any longer. For instance, 
it does not apply to the query  $Q_1$ of the introduction which is 
not nicely-moded.
In fact,
 the leftmost atom of $Q_1$ is resolvable only after the rightmost one has
been resolved at least once. 
%\\

The following immediate corollary will be used in the sequel.

\begin{corollary}
\label{cor-sw}
Let the program $P$ and the query $Q:=\mathbf{ A},\mathbf{ B}$ be nicely-moded.
Suppose that 
%there exists
\[\delta :=\mathbf{ A},\mathbf{ B}\stackrel{\theta}
\longmapsto \mathbf{ C}_1,\mathbf{ C}_2\]
that is a partial input-consuming derivation of $P\U\{Q\}$
where $\mathbf{ C}_1$ and $\mathbf{ C}_2$ are obtained by partially resolving 
$\mathbf{ A}$ and $\mathbf{ B}$, respectively.
Then there exists a partial input-consuming derivation
\[\delta' :=\mathbf{ A},\mathbf{ B}\stackrel{\theta_1}
\longmapsto \mathbf{ C}_1, \mathbf{ B}\theta_1 \stackrel{\theta_2}
\longmapsto \mathbf{ C}_1,\mathbf{ C}_2\]
where 
all the $\mathbf{ A}$-steps are performed in  the prefix 
$\mathbf{ A},\mathbf{ B}\stackrel{\theta_1}
\longmapsto \mathbf{ C}_1, \mathbf{ B}\theta_1 $
%of $\delta'$ 
and $\theta=\theta_1\theta_2$.
\end{corollary}

%%%%%%%%%%%%%%%%%%%%%%%%%%%%%%%%%%%%%%%%%%%%%%%%%%%%%%%%%%%%%%%%%%
\section{Termination}
\label{input-ter}
%%%%%%%%%%%%%%%%%%%%%%%%%%%%%%%%%%%%%%%%%%%%%%%%%%%%%%%%%%%%%%%%%%%

In this section we study the termination of input-consuming
derivations.  To this end we refine the ideas of
Bezem \cite{Bez93} and Cavedon \cite{Cav89} who studied the
termination of logic programs in a very strong sense, namely with
respect to all selection rules, and of Smaus \cite{Sma99-ICLP} who
characterized terminating input-consuming derivations of programs
which are both well and nicely-moded.

\subsection{Input Terminating Programs}
\noindent
We first introduce the key notion of this section.

\begin{definition}[Input Termination]
  A program is called \emph{input terminating} iff all its
  input-consuming derivations started in a nicely-moded query are
  finite.
\end{definition}

The method we 
%are going to
 use in order to prove that a program is
input terminating is based on the following concept of moded level
mapping due to Etalle \emph{et al.} \cite{EBC99}.

\begin{definition}[Moded Level Mapping]
  Let $P$ be a program and ${\cal B}_P^{\cal E}$ be the extended
  Herbrand base\footnote{The extended Herbrand base of $P$ is the set of 
equivalence classes of
all (possibly non-ground) atoms, modulo renaming,  whose
predicate symbol appears in $P$. As usual, an atom is identified with
its equivalence class.} for the
  language associated with $P$.  A function $|\;|$ is a \emph{moded
    level mapping for $P$} iff:
\begin{itemize}
\item it is a function $|\;|:{\cal B}_P^{\cal E} \rightarrow
  \mathbf{N}$ from atoms to natural numbers;
\item for any $\mathbf{ t}$ and $\mathbf{ u}$, $|p(\mathbf{
    s},\mathbf{ t})|=|p(\mathbf{ s},\mathbf{ u})|$.
\end{itemize}
For $A\in {\cal B}_P^{\cal E}$, $|A|$ is the \emph{level} of $A$.
\end{definition}

The condition $|p(\mathbf{ s},\mathbf{ t})|=|p(\mathbf{ s},\mathbf{
  u})|$ states that the level of an atom is independent from the terms
in its output positions. There is actually a small yet important
difference between this definition and the one in \cite{EBC99}: In
\cite{EBC99} the level mapping is defined on ground atoms only.
Indeed, in \cite{EBC99} only well-moded atoms are considered, i.e.,
atoms with ground terms in the input positions.
Here, instead, we are considering nicely-moded atoms whose
input positions can be filled in by (possibly) non-ground terms.

\begin{example}
\label{exa:modedlevelmapping}
  Let us denote by $\mathit{ TSize}(t)$ the term size of a term $t$, that
  is the number of function and constant symbols that occur in $t$.  
\begin{itemize}
\item A moded level mapping for the program {\tt APPEND}
reported in the introduction is as follows:

\begin{program}
\>    |app($\mathit{x_s}$,$\mathit{y_s}$,$\mathit{z_s}$)|=\textrm{\emph{TSize}}($\mathit{x_s}$).
\end{program}

\item A moded level mapping for the program {\tt REVERSE} with accumulator
  of  Example \ref{i-c} is the following:
  
\begin{program}
\>    |reverse($\mathit{x_s}$,$\mathit{y_s}$)|= \textrm{\emph{TSize}}($\mathit{x_s}$)\\
\>    |reverse\_acc($\mathit{x_s}$,$\mathit{y_s}$,$\mathit{z_s}$)|=\textrm{\emph{TSize}}($\mathit{x_s}$).
\end{program}
  
\end{itemize}
\end{example}

%%%%%%%%%%%%%%%%%%%%%%%%
\subsection{Quasi Recurrency}
\noindent
%%%%%%%%%%%%%%%%

In order to give a sufficient condition for termination, we are going
to employ a generalization of the concept of \emph{recurrent} and of
\emph{semi-recurrent} program.  The first notion (which in the case of
normal programs, i.e., programs with negation,
 coincides with the one of \emph{acyclic program}) was
introduced in \cite{Bez93,AB91} and independently in
 \cite{Cav91} in order to prove universal termination for
all selection rules together with other properties of logic
programs. Later, Apt and Pedreschi \cite{AP94} provided the new
definition of semi-recurrent program, which is equivalent to the one
of recurrent program, but it is easier to verify in an automatic
fashion.
In order to proceed, we need a preliminary definition.

\begin{definition} Let $P$ be a program, $p$ and $q$ be relations.
  We say that \emph{$p$ refers to $q$} in $P$ iff there is a clause in
  $P$ with $p$ in the head and $q$ in the body. We say that \emph{$p$
    depends on $q$} and write $p\sqsupseteq q$ in $P$ iff $(p,q)$ is
  in the reflexive and transitive closure of the relation \emph{refers
    to}.
\end{definition}

According to the above definition, $p\simeq q \equiv p \sqsubseteq q
\wedge p \sqsupseteq q$ means that $p$ and $q$ are mutually recursive,
and $p\sqsupset q \equiv p\sqsupseteq q \wedge p\not \simeq q$ means
that $p$ calls $q$ as a subprogram.  Notice that $\sqsupset$ is a
well-founded ordering.

Finally, we can provide the key concept we are going to use in order
to prove input termination.

\begin{definition}[Quasi Recurrency]
  Let $P$ be a program and $|\;|\!:\!{\cal B}_P^{\cal E}
  \rightarrow~\mathbf{ N}$ be a moded level mapping.
\begin{itemize}
\itemsep 0pt
\item[$\bullet$] A clause of $P$ is called \emph{quasi recurrent with
    respect to $|\;|$ } if for every instance of it, $H\leftarrow
  \mathbf{ A},B,\mathbf{ C}$
\begin{equation}
\label{eq:qr}
\mbox{if }\mathit{Rel}(H)\simeq \mathit{ Rel}(B) \mbox{ then }
|H|>|B|.
\end{equation}
\item[$\bullet$] A program $P$ is called \emph{quasi recurrent with
    respect to $|\;|$ } if all its clauses are.  $P$ is called
  \emph{quasi recurrent} if it is quasi recurrent wrt.\ 
  some moded level mapping $|\;|:{\cal B}_P^{\cal E} \rightarrow
  \mathbf{ N}$.
\end{itemize}
\end{definition}

The notion of quasi recurrent program differs from the concepts of
recurrent and of semi-recurrent program in two ways. First, we require that
$|H| > |B|$ only for those body atoms which mutually depend on
$\mathit{Rel}(H)$; in contrast, both the concept of recurrent and
of semi-recurrent program require that $|H| > |B|$ ($|H| \geq |B|$ in the
case of semi-recurrency) also for the atoms for which $\mathit{
  Rel}(H)\not \simeq \mathit{ Rel}(B)$.
Secondly, every instance of a program clause is considered, not only 
ground instances as in the case of (semi-)recurrent programs.
This allows us to treat directly any nicely-moded query 
without introducing the concept of \emph{boundedness} \cite{AP94}
or \emph{cover} as in \cite{MT95}.

It is worthwhile noticing that this concept almost coincides with the
one of \emph{ICD-acceptable} program introduced and used in \cite{Sma99-ICLP}. We
decided to use a different name because we believe that
referring to the word acceptable might lead to confusion: The concept
of acceptable program was introduced by Apt and Pedreschi
\cite{AP93,AP94} in order to prove termination of logic programs using
the  left-to-right selection rule. The crucial difference between
recurrency and acceptability lies in the fact that the latter relies
on  a model $M$; this allows condition (\ref{eq:qr}) to
be checked only for those body atoms which are in a way ``reachable''
wrt.\ $M$. Hence, every recurrent program is acceptable but not
vice-versa.  As an aside, Marchiori and Teusink  \cite{MT95}
introduce the notion of 
\emph{delay recurrent} program 
although
 their concept
is based on the presence of a model $M$.
Our definition does not rely on 
 a model, and so it is  much more related to the notion of recurrent
than the one of acceptable program.

We can  now state our first basic result on termination, in the case of
non-modular programs.

\begin{theorem}
\label{thm:sufficiency}
Let $P$ be a nicely-moded program.  If $P$ is quasi recurrent then $P$
is input terminating.
\end{theorem}
\begin{proof}
It will be obtained from the proof of Theorem \ref{ter-modular}
by setting $R=\emptyset$.
\end{proof}

\begin{example}
  Consider the program $\mathtt{MERGE}$ defined in  Example
  \ref{programs}.  Let $|\;|$ be the moded level mapping for $\mathtt{
    MERGE}$ defined by
\begin{program}
\>  |merge($\mathit{x_s}$,$\mathit{y_s}$,$\mathit{z_s}$)| = \textrm{\emph{TSize}}($\mathit{x_s}$) + \textrm{\emph{TSize}}($\mathit{y_s}$).
\end{program}
It is easy to prove that {\tt MERGE} is quasi recurrent wrt.\ 
 the moded level mapping above. By Theorem \ref{thm:sufficiency},
all input-consuming derivations of  {\tt MERGE} started
with a query $\mathtt{merge}(s,t,u)$,
 where $ u$ is linear and variable disjoint from $ s$
and $t$, are terminating.
\end{example}

%%%%%%%%%%%%%%%%%%%%%%%%%%%%%%%%
\subsection{Modular Termination}
\label{mod}
%%%%%%%%%%%%%%%%%%%%%%%%%%%%%%%%%%

This section contains a generalization of Theorem \ref{thm:sufficiency} to the
modular case, as well as the complete proofs for it.  
The following lemma is a crucial one.

\begin{lemma}
\label{inf}
Let the program $P$ and the query $Q:=A_1,\ldots,A_n$ be nicely-moded.
Suppose that there exists an infinite input-consuming derivation
$\delta$ of $P\U\{Q\}$. Then, there exist an index
$i\in\{1,\ldots,n\}$ and substitution $\theta$ such that
\begin{enumerate}
\item there exists an  input-consuming
derivation $\delta'$ of $P\U\{Q\}$ of the form
\[\delta' := A_1,\ldots,A_n\stackrel{\theta}
\longmapsto \mathbf{ C}, (A_i,\ldots,A_n)\theta \longmapsto \cdots\]
\item there exists an infinite input-consuming derivation
of
$P\U\{A_i\theta\}$.
\end{enumerate}
\end{lemma}
\begin{proof}
Let $\delta:= A_1,\ldots,A_n
\longmapsto \cdots$ be an infinite input-consuming derivation of
$P\U\{Q\}$. Then 
$\delta$ contains an infinite number of $A_k$-steps
for some  $k\in \{1,\ldots,n\}$.
Let $i$ be the minimum of such $k$.
Hence $\delta$ contains a finite number of $A_j$-steps for
$j\in\{1,\ldots,i-1\}$ and there exists $\mathbf{ C}$ and $\mathbf{ D}$
such that 
\[\delta:=A_1,\ldots,A_n \stackrel{\vartheta}
\longmapsto \mathbf{ C},\mathbf{ D}\longmapsto \cdots\] where
$A_1,\ldots,A_n \stackrel{\vartheta} \longmapsto \mathbf{ C},\mathbf{
  D}$ is a finite prefix of $\delta$ which comprises all the
$A_j$-steps of $\delta$ for $j\in\{1,\ldots,i-1\}$ and $\mathbf{C}$ is
the subquery of $\mathbf{C},\mathbf{D}$ consisting of the atoms
resulting from some $A_j$-step ($j\in\{1,\ldots,i-1\}$).
By Corollary~\ref{cor-sw}, there exists an infinite
input-consuming derivation $\delta'$ such that
\[\delta':=A_1,\ldots,A_n \stackrel{\theta} \longmapsto \mathbf{ C},
(A_i,\ldots,A_n)\theta \stackrel{\theta'} \longmapsto \mathbf{ C},
\mathbf{ D} \longmapsto \cdots\] where $\vartheta=\theta\theta'$.
This proves (i).

Now, let $\delta'':=\mathbf{ C}, (A_i,\ldots,A_n)\theta
\stackrel{\theta'} \longmapsto \mathbf{ C}, \mathbf{ D} \longmapsto
\cdots$. Note that in $\delta''$ the atoms of $\mathbf{ C}$ will never
be selected
and, by Remark \ref{Smaus2}, will never be instantiated.  
Let  $\delta'''$ be obtained from  $\delta''$ by omitting the
prefix $\mathbf{ C}$ in each query.
Hence $\delta'''$
is an infinite input-consuming derivation of
$P\U\{(A_i,\ldots,A_n)\theta\}$ where an infinite number of
$A_i\theta$-steps are performed. Again, By Remark \ref{Smaus2}, for
every finite prefix of $\delta'''$ of the form
\[A_i\theta,(A_{i+1},\ldots,A_n)\theta \stackrel{\sigma_1}
\longmapsto  \mathbf{ D}_1,\mathbf{ D}_2 \stackrel{\sigma_2}
\Longrightarrow \mathbf{ D}'_1,\mathbf{ D}'_2\]
where  $\mathbf{ D}_1$ and $\mathbf{ D}_2$ are obtained by partially resolving 
$ A_i\theta$ and $(A_{i+1},\ldots,A_n)\theta$, respectively, and
$\mathbf{ D}_1,\mathbf{ D}_2 \stackrel{\sigma_2}
\Longrightarrow \mathbf{ D}'_1,\mathbf{ D}'_2$
is an $A_j$-step for some
$j\in\{i+1,\ldots,n\}$, we have that $\mathbf{ D}'_1=\mathbf{ D}_1$.
Hence, from the hypothesis that there is an infinite number of 
$A_i\theta$-steps in $\delta''$, it follows that there exists an infinite
input-consuming derivation of $P\U\{A_i\theta\}$. This proves (ii).
\end{proof}
\vspace{3mm}

The importance of the above lemma is shown by the following corollary
of it, which will allow us to concentrate  on queries
containing only one atom.

\begin{corollary}
\label{cor:inf-one-atom}
Let $P$ be a nicely-moded program. $P$ is input terminating iff for
each nicely-moded one-atom query $A$ all input-consuming derivations
of $P\U\{A\}$ are finite.
\end{corollary}

\noindent
We can now state the main result of this section. Here and in what
follows we say that a relation $p$ is \emph{defined in} the program
$P$ if $p$ occurs in a head of a clause of $P$, and that $P$
\emph{extends} the program $R$ if no relation defined in $P$ occurs
in $R$.

\begin{theorem}
\label{ter-modular}
Let $P$ and $R$ be two programs such that $P$ extends $R$. 
 Suppose that 
\begin{itemize}
\item $R$ is input terminating,
\item $P$ is nicely-moded and quasi recurrent wrt.\ 
 a moded level
mapping $|\;|:{\cal B}_P^{\cal E}
\rightarrow~\mathbf{ N}$.
\end{itemize}
Then $P\U R$ is input terminating.
\end{theorem}
\begin{proof}
  First, for each predicate symbol $p$, we define $\mathit{ dep}_P(p)$ to
  be the number of predicate symbols it depends on. More formally,
  $\mathit{ dep}_P(p)$ is defined as the cardinality of the set $\{q|\; q
  \mbox{ is defined in } P \mbox{ and } p\sqsupseteq q\}$.  Clearly,
  $\mathit{ dep}_P(p)$ is always finite. Further, it is immediate to see
  that if $p\simeq q$ then $\mathit{ dep}_P(p)=\mathit{ dep}_P(q)$ and that if
  $p\sqsupset q$ then $\mathit{ dep}_P(p)>\mathit{ dep}_P(q)$.
  
  We can now prove our theorem. By Corollary \ref{cor:inf-one-atom},
  it is sufficient to prove that for any nicely-moded one-atom query
  $A$, all input-consuming derivations of $P\U\{A\}$ are finite.

  First notice that if $A$ is defined in $R$ then the result follows
  immediately from the hypothesis that $R$ is input terminating and
  that $P$ is an extension of $R$. So we can assume that $A$ is
  defined in $P$.
  
For the purpose of deriving a contradiction, assume that
 $\delta$ is an infinite input-consuming derivation of $(P\U R)
  \U \{A\}$ such that $A$ is defined in $P$. Then
\[\delta:=A\stackrel{\theta_1}\Longrightarrow (B_1,\ldots,B_n)\theta_1
\stackrel{\theta_2} \Longrightarrow \cdots \]
where $H\leftarrow B_1,\ldots,B_n $ is the
input clause used in the first derivation step and 
$\theta_1=
\mathit{ mgu}(A , H)$. Clearly,
$(B_1,\ldots,B_n)\theta_1$ has an infinite 
input-consuming derivation in $P\U R$.
By Lemma \ref{inf}, 
for some $i\in\{1,\ldots,n\}$ and for some substitution $\theta_2$,
\begin{enumerate}
\item
there exists 
an infinite input-consuming 
derivation of $(P\U R)\U \{ A\}$ of the form
\[A\stackrel{\theta_1}\Longrightarrow (B_1,\ldots,B_n)\theta_1
\stackrel{\theta_2} \longmapsto \mathbf{ C}, (B_i,\ldots,B_n)
\theta_1\theta_2\cdots ;\]
\item there exists an infinite input-consuming derivation
of $P\U \{B_i\theta_1\theta_2\}.$
\end{enumerate}
Notice also that $B_i\theta_1\theta_2$ is nicely-moded.
Let now $\theta=\theta_1\theta_2$.  Note that $H\theta\leftarrow
(B_1,\ldots,B_n)\theta $ is an instance of a clause of~$P$.

We  show that (2) cannot hold. This is done by induction on
$\langle \mathit{ dep}_P(\mathit{ Rel}(A)), |A|\rangle$ wrt.\ 
 the ordering $\succ$ defined by: $\langle m,n\rangle \succ \langle
m',n'\rangle$ iff either $m>m'$ or $m=m'$ and $n>n'$.

{\it Base}.  Let $\mathit{ dep}_P(\mathit{Rel}(A))=0$ ($|A|$ is arbitrary).
 In
this case, $A$ does not depend on any predicate symbol of $P$, thus
all the $B_i$ as well as all the atoms occurring in its descendents in
any input-consuming derivation are defined in $R$. The hypothesis that
$R$ is input terminating contradicts $(2)$ above.

\emph{Induction step}. 
We distinguish two cases:
\begin{enumerate}
\itemsep 0pt
\item $\mathit{ Rel}(H)\sqsupset \mathit{ Rel}(B_i)$,
\item $\mathit{ Rel}(H)\simeq \mathit{ Rel}(B_i)$.
\end{enumerate}
In case $(a)$ we have that 
$\mathit{ dep}_P(\mathit{ Rel}(A))
=\mathit{ dep}_P(\mathit{ Rel}(H\theta))>
\mathit{ dep}_P(\mathit{ Rel}(B_i\theta))$. So, 
$\langle \mathit{ dep}_P(\mathit{ Rel}(A)), |A|\rangle=
\langle \mathit{ dep}_P(\mathit{ Rel}(H\theta)), |H\theta|\rangle \succ
\langle \mathit{ dep}_P(\mathit{ Rel}(B_i\theta)), |B_i\theta|\rangle$.
In case~$(b)$,
from the hypothesis
 that $P$ is quasi recurrent wrt.\ $|\;|$, it follows that
$|H\theta|> |B_i\theta|$.

Consider now the partial input-consuming derivation $A\stackrel{\theta}
\longmapsto \mathbf{ C}, (B_i,\ldots,B_n) \theta $.
 By Corollary~\ref{corollary} 
and the fact that $|\;|$ is a moded level mapping, it follows that
 $|A|=|A\theta|=|H\theta|$.  Therefore, $\langle \mathit{
  dep}_P(\mathit{ Rel}(A)), |A|\rangle =\langle \mathit{
  dep}_P(\mathit{ Rel}(H\theta)), |H\theta|\rangle \succ \langle
\mathit{ dep}_P(\mathit{ Rel}(B_i\theta)), |B_i\theta|\rangle$.  In
both cases, the contradiction follows by the inductive hypothesis.
\end{proof}

\begin{example}
\label{exa:flatten}
The program {\tt FLATTEN} using difference-lists is nicely-moded with
respect to
the modes described below, provided that one replaces ``$\setminus$''
by ``,'', as we have done here.
\begin{program}
\>  mode   flatten(In,Out).\\
\>  mode   flatten\_dl(In,Out,In).\\
\>  mode   constant(In).\\
\>  mode   $\neq$(In,In).\\[2mm]
\>  flatten(Xs,Ys) \la  flatten\_dl(Xs,Ys,[ ]).\\[2mm]
\>  flatten\_dl([ ],Ys,Ys).\\
\>  flatten\_dl(X,[X|Xs],Xs) \la  constant(X),
 X $\neq$ [ ].\\
\> flatten\_dl([X|Xs],Ys,Zs) \la \= flatten\_dl(Xs,Y1s,Zs), \\
\> \> flatten\_dl(X,Ys,Y1s).
\end{program}
Consider the moded level mapping for  $\mathtt{ FLATTEN}$
defined by
\begin{program}
\>    |flatten($\mathit{x_s}$,$\mathit{y_s}$)|  = \textrm{$\emph{TSize}$}($\mathit{x_s}$)\\
\>    |flatten\_dl($\mathit{x_s}$,$\mathit{y_s}$,$\mathit{z_s}$)|  =  \textrm{$\emph{TSize}$}($\mathit{x_s}$).
\end{program}
It is easy to see that the program {\tt FLATTEN} is quasi recurrent
wrt.\ the moded level mapping above. Hence, all
input-consuming derivations of program {\tt FLATTEN} started with a query
\texttt{flatten($s$,$t$)},  where $t$
is linear and variable disjoint from $s$, are terminating.
\end{example}

%%%%%%%%%%%%%%%%%%%%%%%
\section{Termination:~A  Necessary Condition}
\label{sec:necessity}
%%%%%%%%%%%%%%%%%%%%%%%%%%%

Theorem \ref{thm:sufficiency} provides a sufficient condition for
termination. The condition is not necessary, as demonstrated by the
following simple example.
\begin{program}
  \> mode p(In,Out).\\[2mm]
  \> p(X,a) \la p(X,b).\\
  \> p(X,b).
\end{program}
This program is clearly input terminating, however it is not quasi
recurrent. If it was, we would have that $|\mathtt{ p(X,a)}|>|\mathtt{
  p(X,b)}|$, for some moded level mapping $|\;|$ (otherwise the first
clause would not be quasi recurrent).  On the other hand, since
$\mathtt{ p(X,a)}$ and $\mathtt{ p(X,b)}$ differ only for the terms
filling in their output positions, by definition of moded level
mapping, $|\mathtt{ p(X,a)}| = |\mathtt{ p(X,b)}|$. Hence, we
have a contradiction.

Nevertheless, as shown by other works, e.g., \cite{Bez93,AP93,EBC99},
it is important to be able to give a characterization of termination,
i.e., a condition which is \emph{necessary} and sufficient to ensure
termination. To this purpose is dedicated this section.

\subsection{Simply-Moded Programs}
As demonstrated by the example above, in order to 
provide a necessary condition for termination
we need to further restrict the
class of programs we consider. The first problem is that we should
rule out those situations in which termination is guaranteed by the
instantiation of the output positions of some selected atom, as it
happens in the above example. For this we restrict to
\emph{simply-moded} programs which are nicely-moded programs with the
additional condition that the output arguments of clause bodies are
variables.

\begin{definition}[Simply-Moded]
\mbox{}
\begin{itemize}
\item
A query $Q$ (resp., a  clause $c=H\la Q$)
is \emph{simply-moded} if it is  nicely-moded
and $\mathit{Out}(Q)$ is a linear sequence of variables.
\item A program $P$ is simply-moded iff all of its clauses are simply-moded.
\end{itemize}
\end{definition}

It is important to notice that most programs are simply-moded (see the
mini-survey at the end of \cite{AP93}) and that often non simply-moded
programs can naturally be transformed into simply-moded ones.

\begin{example}
\label{programs-sm}
\mbox{}
\begin{itemize}
\item The programs {\tt REVERSE} of Example \ref{i-c}, {\tt MERGE} of
  Example \ref{programs} and {\tt FLATTEN} of Example
  \ref{exa:flatten} are all simply-moded.  
\item Consider the program {\tt LAST} which extends {\tt
    REVERSE}:
\begin{program}
  \> mode last(In,Out).\\
  \> last(Ls,E)\la reverse(Ls,[E|\_]).
\end{program}
This program is not simply-moded since the argument filling in the
output position in the body of the first clause is not a variable.
However, it can be transformed into a simply-moded one as follows:
\begin{program}
\> mode last(In,Out).\\
\> mode selectfirst(In,Out).\\[2mm]
\> last(Ls,E)\la reverse(Ls,Rs), selectfirst(Rs,E).\\
\> selectfirst([E|\_],E).
\end{program}
\end{itemize}
\end{example}
The following lemma, which is an immediate consequence of Lemma 30 in
\cite{AL95}, 
shows the persistence of the notion of simply-modedness.
\begin{lemma}
\label{lemma:s-persistence}
Every resolvent of a simply-moded query $Q$ and a simply-moded clause
$c$, where the derivation step is input-consuming and $\mathit{
  Var}(Q)\cap \mathit{ Var}(c)=\emptyset$, is simply-moded.  
\end{lemma}

\subsection{Input-Recursive Programs}

Unfortunately, the restriction to simply-moded programs alone is not
sufficient to extend Theorem \ref{thm:sufficiency} by
 a necessary condition. 
Consider for instance the following
program \texttt{QUICKSORT}:

\begin{program}
\> \>mode qs(In,Out).\\
\> \>mode part(In,In,Out,Out).\\
\> \>mode app(In,In,Out).\\[2mm]
\> \>qs([ ],[ ]).\\
$c_1\!:=$ \> \>qs([X|Xs],Ys) \la part(X,Xs,Littles,Bigs),\\
\> \> \>qs(Littles,Ls),\\
\> \> \>qs(Bigs,Bs), \\
\> \> \>app(Ls,[X|Bs],Ys).\\[2mm]
\>  \>part(X,[ ],[ ],[ ]).\\
\>  \>part(X,[Y|Xs],[Y|Ls],Bs) \la X>Y, part(X,Xs,Ls,Bs).\\
\>  \>part(X,[Y|Xs],Ls,[Y|Bs]) \la X<=Y, part(X,Xs,Ls,Bs).
\end{program}

This program is simply-moded and input terminating\footnote{Provided
  that one models the built-in predicates \texttt{>} and \texttt{<=}
  as being defined by (an infinite number of) ground facts of the form
  \texttt{>($m$,$n$)} and \texttt{<=($m$,$n$)}. 
The problem here is that the definition of
  input-consuming derivation does not consider the presence of
  built-ins.}. However it is not quasi recurrent. Indeed, there exist
no moded level mapping $|\;|$ such that, for every variable-instance,
$|\mathtt{qs([X|Xs],Ys)}|> |\mathtt{qs(Littles,Ls)}|$ and
$|\mathtt{qs([X|Xs],Ys)}|> |\mathtt{qs(Bigs,Bs)}|$.  This is due to
the fact that, in clause $c_1$ there is no direct link between
the input arguments of the recursive calls and those of the clause
head. This motivates the following definition of
 \emph{input-recursive} programs.

\begin{definition}[Input-Recursive]
Let $P$ be a program.
\begin{itemize}
\itemsep 0pt
\item A clause  $H\leftarrow
  \mathbf{ A},B,\mathbf{ C}$ of $P$ is called \emph{input-recursive} if
\[\mbox{if }\mathit{Rel}(H)\simeq \mathit{ Rel}(B) \mbox{ then }
\mathit{Var}(\mathit{In}(B))\subseteq \mathit{Var}(\mathit{In}(H)).\]
\item A program $P$ is called \emph{input-recursive}
 if all its clauses are.
\end{itemize}
\end{definition}

Thus, we say that a clause is input-recursive if the set of variables
occurring in the arguments filling in the input positions of each
recursive call in the clause body is a subset of the set of variables
occurring in the arguments filling in the input positions of the
clause head. Input-recursive programs have strong similarities with
primitive recursive functions.

\begin{example}
\label{programs-ir}
\mbox{}
\begin{itemize}
\item The programs {\tt APPEND} of the introduction, {\tt REVERSE} of
  Example \ref{i-c} and {\tt MERGE} of Example \ref{programs} are all
  input-recursive.
  
\item The program {\tt FLATTEN} of Example \ref{exa:flatten} is not
  input-recursive. This is due to the presence of the fresh variable
  \texttt{Y1s} in a body atom of the last clause.
  
\item \texttt{QUICKSORT}, is not input-recursive. In particular,
  clause $c_1$ is not input-recursive.
\end{itemize}
\end{example}

\subsection{Characterizing Input Terminating Programs}

We can now prove that by restricting ourselves to input-recursive and
simply-moded programs, the condition of Theorem \ref{thm:sufficiency}
is also a necessary one.

To prove this, we follow the approach of Apt and Pedreschi when
characterizing terminating programs \cite{AP94}. First we introduce
the notion of \emph{IC-tree} that corresponds to the notion of
\emph{S-tree} in \cite{AP94} and provides us with a representation for
all input-consuming derivations of a program $P$ with a query $Q$,
then we define a level mapping which associates to every atom $A$ the
number of nodes of a given IC-tree and finally we prove that $P$ is
quasi recurrent wrt.\  such a level mapping.

\begin{definition}[IC-tree]
\label{def:i-tree}
An \textit{IC-tree} for $P\cup\{Q\}$ is a tree whose nodes are 
labelled with queries such that
\begin{itemize}
\item its branches are input-consuming derivations of
 $P\cup\{Q\}$,
\item every node $Q$ has exactly one descendant for every atom $A$
of $Q$ and every clause $c$ from $P$ such that $A$ is input-consuming
resolvable wrt. $c$. This descendant is a resolvent of $Q$ and $c$ wrt. $A$.
\end{itemize}
\end{definition}
In this tree, a node's children consist of all its resolvents, ``modulo
renaming'', via an input-consuming derivation step wrt.\ 
 all the possible
choices of a program clause and a selected atom.  
\begin{lemma}[IC-tree 1]
\label{lemma:i-tree1}
An IC-tree for $P\cup\{Q\}$ is finite iff all input-consuming
derivations of
$P\cup\{Q\}$ are finite.
\end{lemma}
\begin{proof}
By definition, the IC-trees are finitely branching.
The claim now follows by K\"{o}nig's Lemma.
\end{proof}
\vspace{3mm}

Notice that if  an IC-tree for $P\cup\{Q\}$ is finite then all
the  IC-trees for $P\cup\{Q\}$ are finite. 

For a program $P$ and a query $Q$, we denote by
 $\mathit{nodes}^\mathit{ic}_P(Q)$
the number of nodes in an IC-tree for $P\cup\{Q\}$.
The following properties of IC-trees will be needed.

\begin{lemma}[IC-tree 2]
\label{lemma:i-tree2}
Let $P$ be a program, $Q$ be a query and $T$ be a finite
 IC-tree for $P\cup\{Q\}$. Then
\begin{itemize}
\item[(i)]  for all non-root nodes $Q'$ in $T$,
   $\mathit{nodes}^\mathit{ic}_P(Q')<
 \mathit{nodes}^\mathit{ic}_P(Q)$,
\item[(ii)] for all atoms $A$ of $Q$,  $\mathit{nodes}^\mathit{ic}_P(A)\leq
 \mathit{nodes}^\mathit{ic}_P(Q)$.
\end{itemize}
\end{lemma}
\begin{proof}
Immediate by Definition \ref{def:i-tree} of IC-tree.
\end{proof}
\vspace{3mm}

We can now prove the desired result.

\begin{theorem}
\label{thm:necessity}
Let $P$ be a simply-moded and input-recursive program.  If $P$ is
input terminating then $P$ is quasi recurrent.
\end{theorem}
\begin{proof}
We show that there exists a moded level mapping $|\;|$ for $P$ such that
$P$ is quasi recurrent wrt.\ $|\;|$.

Given an atom $A$, we denote with $A^*$ an atom obtained from $A$
by replacing the terms filling in its output positions with fresh
distinct variables. Clearly, we have that $A^*$ is simply-moded.
Then we define the following moded level mapping 
for $P$:
\[ |A| = \mathit{nodes}^\mathit{ic}_P(A^*). \]
Notice that, the level $|A|$
of an atom $A$ is independent from the terms filling in its output 
 positions, i.e., $|\;|$ is a moded level mapping. Moreover, since $P$ is 
input terminating and $A^*$ is simply-moded (in particular, it is
nicely-moded), all the input-consuming derivations of 
$P\cup\{A^*\}$ are finite.
Therefore, 
by Lemma~\ref{lemma:i-tree1}, 
$\mathit{nodes}^\mathit{ic}_P(A^*)$  is defined (and finite), and thus
$|A|$ is defined (and finite) for every atom $A$.

We now prove that $P$ is quasi recurrent wrt.\  $|\;|$.

Let  $c: H\leftarrow \mathbf{ A},B,\mathbf{ C}$ be a clause of $P$
and  $ H\theta\leftarrow \mathbf{ A}\theta,B\theta,\mathbf{ C}\theta$
be an instance of $c$ (for some substitution $\theta$).
We show that 
$\mbox{if }\mathit{Rel}(H)\simeq \mathit{ Rel}(B) \mbox{ then }
|H\theta|>|B\theta|$.

Let $H=p(\ol s,\ol t)$. Hence, $(H\theta)^*=p(\ol s\theta,\ol x)$
where $\ol x$ is a sequence of fresh distinct variables.  Consider a
variant $c': H'\leftarrow \mathbf{ A}',B',\mathbf{ C}'$ of $c$
variable disjoint from $(H\theta)^*$.  Let $\rho$ be a renaming such
that $c'=c\rho$.  Clearly, $(H\theta)^*$ and $H'$ unify.  Let
$\mu=\mathit{mgu}((H\theta)^*, H')=\mathit{mgu}((H\theta)^*, H\rho)=
\mathit{mgu}(p(\ol s\theta,\ol x), p(\ol s,\ol t)\rho)$.  By
properties of substitutions (see \cite{Apt97}), since $\ol x$ consists
of fresh variables, there exists two relevant mgu $\sigma_1$ and
$\sigma_2$ such that
\begin{itemize}
\item $\sigma_1=\textit{mgu}(\ol s\theta, \ol s\rho)$,
\item $\sigma_2=\textit{mgu}(\ol x\sigma_1, \ol t\rho\sigma_1)$.
\end{itemize}
Since $\ol s\rho \leq \ol s\theta$,
we can assume that $\mathit{Dom}(\sigma_1)\subseteq \mathit{Var}(\ol s\rho)$.
 Because of standardi\-zation apart,
since $\ol x$ consists of fresh variables, $\ol x\sigma_1=\ol x$ and
thus $\sigma_2=\textit{mgu}(\ol x, \ol t\rho\sigma_1)$.
Since $\ol x$
is a sequence of variables, we can also assume that
  $\mathit{Dom}(\sigma_2)\subseteq
\mathit{Var}(\ol x)$. Therefore $\mathit{Dom}(\mu)\subseteq
\mathit{Var}(\mathit{Out}((H\theta)^*))\cup
\mathit{Var}(\mathit{In}(H\rho))$. Moreover, since $(\mathbf{
  A}',B',\mathbf{ C}')\mu=(\mathbf{ A},B,\mathbf{ C})\rho\mu$, we have
that
$$(H\theta)^*\stackrel{\mu}\Longrightarrow (\mathbf{ A},B,\mathbf{ C})\rho\mu$$
is an input-consuming derivation step, i.e.,
$(\mathbf{ A},B,\mathbf{ C})\rho\mu$
is a descendant of $(H\theta)^*$ in an IC-tree for $P\cup\{(H\theta)^*\}$.

By definition of $\mu$,
 $\ol s\theta = \ol s\rho\mu$; hence
\begin{equation}
\label{eq:stheta}
(\rho\mu)_{|\mathit{In}(H)}=\theta_{|\ol s}.
\end{equation}

Let now $B=p(\ol u,\ol v)$.
By (\ref{eq:stheta}) and the hypothesis  that $c$ is input-recursive, that is
$\mathit{Var}(\mathit{In}(B))\subseteq 
\mathit{Var}(\mathit{In}(H))= \mathit{Var}(\ol s)$, it follows that
\begin{equation}
\label{eq:utheta}
\ol u\rho\mu=\ol u (\rho\mu)_{|\mathit{In}(H)}=\ol u\theta_{|\ol s}=
\ol u\theta.
\end{equation}
Moreover, since $c'$ is simply-moded,
$\mathit{In}(H\rho)\cap \mathit{Out}(B\rho)=\emptyset$.
Hence, by definition of $\mu$
and standardization apart,
 $\mathit{Dom}(\mu)\cap  \mathit{Out}(B\rho)=\emptyset$, i.e.,
\begin{equation}
\label{eq:vrhomu}
\ol v\rho\mu=\ol v\rho.
\end{equation}
 Therefore,
by (\ref{eq:utheta}) and (\ref{eq:vrhomu}), 
$B\rho\mu=p(\ol u,\ol v) \rho\mu= 
 p(\ol u\theta,\ol v\rho) = (B\theta)^*$, i.e.,
\begin{equation}
\label{eq:Btheta*}
B\rho\mu = (B\theta)^*.
\end{equation}
Hence,\\
\\
\(\begin{array}{llll}
|H\theta| & = & \mathit{nodes}^\mathit{ic}_P((H\theta)^*) & 
\mbox{by definition of } |\;|\\
& > & \mathit{nodes}^\mathit{ic}_P( (\mathbf{ A},B,\mathbf{ C})\rho\mu)
& \mbox{by Lemma \ref{lemma:i-tree2} (i)}\\
& \geq & \mathit{nodes}^\mathit{ic}_P(B\rho\mu) & \mbox{by Lemma \ref{lemma:i-tree2} (ii)}\\
& = & \mathit{nodes}^\mathit{ic}_P((B\theta)^*)& \mbox{by (\ref{eq:Btheta*})}\\
& = & |B\theta| & \mbox{by definition of } |\;|.
\end{array}\)\\
\mbox{ }\end{proof}

%%%%%%%%%%%%%%%%%%%%%%%%%%
\section{Applicability}
\label{sec:applicability}
%%%%%%%%%%%%%%%%%%%%%%%%%%%%

This section is intended to 
show through some examples the applicability of our results.
Then, programs from various well-known collections
are analyzed.

\subsection{Examples}
\label{sec:examples}

It is worth noticing that, since the definition of
input-consuming derivation is independent from the textual order of
the atoms in the clause bodies,
 the results we have provided (Theorems \ref{thm:sufficiency},
\ref{ter-modular} and \ref{thm:necessity}) hold also in the case that
programs and queries are \emph{permutation} nicely- (or
simply-) moded \cite{SHK98}, that is programs and
queries which would be nicely- (or simply-) moded after a permutation
of the atoms in the bodies.  
Therefore, for instance, we can apply
Theorems \ref{thm:sufficiency} and \ref{ter-modular} to the program
$\mathtt{ FLATTEN}$ as it is presented in \cite{Apt97} (except for the
replacement of ``$\setminus$'' with ``,''), i.e.,

\begin{program}
\>  flatten(Xs,Ys) \la flatten\_dl(Xs,Ys,[ ]).\\[2mm]
\>  flatten\_dl([ ],Ys,Ys).\\
\>  flatten\_dl(X,[X|Xs],Xs) \la  constant(X),  X $\neq$ [ ].\\
\>  flatten\_dl([X|Xs],Ys,Zs) \la  \= flatten\_dl(X,Ys,Y1s),\\ 
\> \> flatten\_dl(Xs,Y1s,Zs).
\end{program}
where the atoms in the body of the last clause are permuted with
respect to the version of  Example \ref{exa:flatten}.
\\

Let us  consider again 
the program \texttt{APPEND} 
 of  the introduction with its natural delay declaration:

\begin{program}
\> mode app(In,In,Out)\\[2mm]
\>  app([ ],Ys,Ys).\\
\> app([H|Xs],Ys,[H|Zs]) \la app(Xs,Ys,Zs).
\end{program}

\begin{program}
\> delay app(Xs,\_,\_) until nonvar(Xs).
\end{program}

Let $\cal Q$ be the set of one-atom queries of the form
\texttt{app($s$,$t$,$Z$)} where $s$ and $t$ are any terms and
$Z$ is a variable disjoint from $s$ and $t$.
Observe that $\cal Q$ is closed under resolution: Each resolvent in a
derivation starting in a query from $\cal Q$ is still a query from
$\cal Q$. Moreover, because of the presence of the delay declaration,
 only atoms whose first argument is a non-variable term
are allowed to be selected. Thus, selectable atoms have the form
\texttt{app($s$,$t$,$Z$)} where
\begin{itemize}
\item[$(1)$]
$s$ is a non-variable
term,
\item[$(2)$]  $t$ is any
term and $Z$ is a variable disjoint from
 $s$ and $t$.
\end{itemize}

Any derivation of \texttt{APPEND} starting in a query of
$\cal Q$ is similar to an input-consuming one. This follows from
 the fact that for any selectable atom $A$ and clause's head
$H$, there exists a mgu $\theta$ which does not affect the input
arguments of $A$.
In fact, let
 $A$ be a selectable atom of $\cal Q$. 
If $A$ unifies with the head of the first clause
then, by $(1)$, $s$ is the empty list
 \texttt{[ ]}
  and $\theta=\mathit{mgu}(A,H)=\{\mathtt{Ys}/t,Z/t\}$.
Otherwise, If $A$ unifies  with the head of the second clause
 then, by
  $(1)$, $s$ is a term of the form
\texttt{[$s_1$|$s_2$]} and
  $\theta=\mathit{mgu}(A,H)=\{\mathtt{H}/s_1,\mathtt{Xs}/s_2,\mathtt{Ys}/t,Z/[s_1|\mathtt{Zs}]\}$.
By $(2)$ it follows that, in both cases, $s\theta=s$
and $t\theta=t$, i.e., $\theta$ does not affect the
input arguments of $A$.

Moreover, it is easy to check that {\tt APPEND} is quasi recurrent
wrt.\  the moded level mapping depicted in Example
\ref{exa:modedlevelmapping}. Since it is nicely-moded, by applying
Theorem \ref{thm:sufficiency} it follows that it is input terminating.
By the arguments above, we can conclude that all the derivations of
$\mathtt{APPEND}$ in presence of the delay declaration $d_1$ and
starting in a (permutation) nicely-moded query are finite.  Hence, in
particular, we can state that all the derivations of \texttt{APPEND}
starting in the query $Q_1$ of the introduction, which is not
nicely-moded but it is permutation nicely-moded, are finite.

\subsection{Benchmarks} 
\label{sec:benchmarks}
In order to assess the applicability of our results, we have looked
into four collections of logic programs, and we have checked those
programs against the three classes of programs: (permutation)
nicely-moded, input
terminating and quasi recurrent programs. The results
are reported in Tables
1 to 4.  These tables clearly show that our results apply to the
large majority of the programs considered.

In Table 1 the programs from Apt's collection are considered, see
\cite{Apt97}.  The programs from the DPPD's
collection, maintained by Leuschel and available at the URL:
http://dsse.ecs.soton.ac.uk/$\sim$mal/systems/dppd.html, are referred
to in Table 2.  Table 3 considers various programs from
Lindenstrauss's collection (see the URL:
http://www.cs.huji.ac.il/$\sim$naomil).  Finally, in Table 4 one finds
the (almost complete) list of programs by F. Bueno, M. Garcia de la
Banda and M.  Hermenegildo that can be found at the URL:
http://www.clip.dia.fi.upm.es.

For each program we specify the name and the modes of the main
procedure. Then we report whether or not the program is (permutation)
nicely-moded (\textbf{NM}), input terminating (\textbf{IT}), and quasi
recurrent (\textbf{QR}).  Notice that for programs which are not input
terminating, because of Theorem \ref{thm:sufficiency}, it does not make
sense to check whether or not they are quasi recurrent.  For this
reason, we leave blank the cells in the column \textbf{QR}
corresponding to non-input terminating programs.

Finally, Table 5 reports the list of programs from previous tables
which have been found to be input terminating but not quasi recurrent.
For these programs, the notion of quasi recurrency does not provide an
exact characterization of input termination. In particular, Theorem
\ref{thm:necessity} does not apply. In order to understand which of
the hypothesis of the theorem does not hold, we report in Table 5
whether or not these programs are simply-moded (\textbf{SM}) and
input-recursive (\textbf{IR}).

\section{Conclusion and Related Works}
\label{sec:conclusion}

In this paper we studied the properties of input-consuming derivations
of nicely-moded programs.
 
This study is motivated by the widespread use of programs using
dynamic schedu\-ling controlled by delay declarations.  In fact, as we
have motivated in Section \ref{sec:ic-dd}, we believe that in most
practical programs employing delay declarations these constructs are
used for guaranteeing that the derivation steps are input-consuming.

In the first place, we showed that for nicely-moded programs a weak
version of the well-known switching lemma holds: If, given a query
$(\ol A, B, \ol C, D, \ol E)$, $D$ is selected before $B$ in an
input-consuming derivation, then the two resolution steps can be
interchanged while maintaining that the derivation is input-consuming.

Secondly, we presented a method for proving termination of programs
and queries which are (permutation) nicely-moded. We also showed a
result characterizing a class of input terminating programs.

In the literature, the paper most related to the present one is
certainly \cite{Sma99-ICLP}. Our results strictly generalize those in
\cite{Sma99-ICLP} in the fact that we drop the condition that programs and
queries have to be well-moded. This is particularly important in the
formulation of the queries. For instance, in the program
\texttt{FLATTEN} of Example \ref{exa:flatten}, our results show that
every input-consuming derivation starting in a query of the form
$\mathtt{flatten}(t,s)$ terminates provided that $t$ is linear and
disjoint from $s$, while the results of \cite{Sma99-ICLP} apply only if
$t$ is a ground term.  Note that well-moded queries (in well-moded
programs) never terminate by deadlock, since the leftmost atom of each
resolvent is ground in its input positions and hence selectable.  This
does not hold for nicely-moded queries which might deadlock. Our
method allows us thus to cope also with this more difficult situation:
For instance we can prove that all derivations of \texttt{APPEND}
starting in $\mathtt{app(X,Y,Z)}$ are terminating.  In practice the
result of \cite{Sma99-ICLP} identify a class of programs and queries which
is both terminating and \emph{deadlock free}.  While deadlock is
clearly an undesirable situation, there are various reasons why one
might want to prove termination independently from the absence of
deadlock: In the first place, one might want to prove absence of
deadlock using a different tool than by employing well-moded programs.
Secondly, in some situations absence of deadlock might be difficult or
impossible to prove, like in a modular context in which the code of
some module is not known, hence not analyzable: consider for instance
the query \texttt{generator\_1(X1s), generator\_2(X2s),
  append(X1s,X2s,Zs).}, where the generators are defined in different
modules; our results allow us to demonstrate that if the
\texttt{generator}s terminate, then the above query terminates.  On
the other hand, one cannot determine whether it is deadlock free
unless one has a more precise specification of the generators.
Thirdly, it is well-known that one of the goals of dynamic scheduling
is precisely enforcing termination; in this respect a deadlock can be
regarded as the situation in which ``all else failed''. Our system
allows us to check how effective dynamic scheduling is in enforcing
termination.

Concluding our comparison with \cite{Sma99-ICLP}, for the class of
(permutation) simply-moded and input-recursive programs, we provide 
 an exact characterization of input termination.  A similar result
is not present in \cite{Sma99-ICLP}.

 Apt and Luitjes \cite{AL95} have also tackled the problem of the
termination of programs in presence of dynamic scheduling.  The
techniques employed in it are based on determinacy checks and the
presence of successful derivations, thus are completely different from
ours. It is nevertheless worth mentioning that \cite{AL95} reports a
special ad-hoc theorem, in order to prove that, if $u$ is linear and
disjoint from $s$ then the query $\mathtt{app}(s,t,u)$ terminates.
This is reported in order to show the difficulties one encounters in
proving termination in presence of dynamic scheduling.  Now, under the
further (mild) additional condition that $u$ be disjoint from $t$, the
termination of $\mathtt{app}(s,t,u)$ is a direct consequence of our
main result.

Another related paper is the one by Marchiori and Teusink \cite{MT95}.
However,  Marchiori and Teusink make a strong restriction on the
selection rule, which has to be \emph{local}; this restriction
actually forbids any form of coroutining. Moreover, \cite{MT95} allows
only \emph{safe} delay declarations; we do not report here the
definition of \emph{safe} delay declaration, we just say that it is
rather restrictive: For instance, the delay declaration
 we have used
for \texttt{APPEND} is not safe (a safe one would be \texttt{delay
  app(X,\_,\_) until list(X)}).
Actually, their requirements 
 go beyond ensuring that derivations are input-consuming.

 Applicability and effectiveness of our results have been demonstrated
 by matching our main definitions against the programs of four public
 program lists. These benchmarks showed that most of the considered
 programs are nicely-moded (for a suitable mode) and quasi recurrent
 (wrt.\ a suitable level mapping).

%\bibliographystyle{alpha}
%\bibliography{sandro}  

\newpage

\begin{table}
\caption{Programs from Apt's Collection}
\label{1}
\centering
\begin{tabular}{lccc||lccc}
\hline\hline
& {\bf NM} & {\bf IT} &   {\bf  QR} & &  {\bf NM} & {\bf IT} &   {\bf  QR}\\
\cline{1-8}
app(In,\_,\_) & yes  & yes & yes &  ordered(In) & yes & yes & yes \\
\cline{1-8}
app(\_,\_,In) & yes  & yes & yes & overlap(\_,In) & yes & yes & yes\\
\cline{1-8}
app(Out,In,Out) & yes  & no &  & overlap(In,Out) & yes & no &\\
\cline{1-8}
append3(In,In,In,Out) & yes & yes & yes &  perm(\_,In) &  yes & yes & yes \\
\cline{1-8}
color\_map(In,Out) & yes & no &  & perm(In,Out) &  yes & no &\\
\cline{1-8}
color\_map(Out,In) & yes & no &  & qsort(In,\_) & yes & yes & no\\
\cline{1-8}
color\_map(In,In) & yes & yes & yes &  qsort(Out,In) & yes & no &\\
\cline{1-8}
dcsolve(In,\_) & yes & no &  &reverse(In,\_) & yes & yes & yes\\
\cline{1-8}
even(In) & yes & yes & yes &  reverse(Out,In) & yes & no &\\
\cline{1-8}
fold(In,In,Out) & yes & yes & yes & select(\_,In,\_) & yes & yes & yes\\
\cline{1-8}
list(In) & yes & yes & yes &select(\_,\_,In) & yes & yes & yes\\
\cline{1-8}
 lte(In,\_) & yes & yes & yes & select(In,Out,Out) & yes & no & \\
\cline{1-8}
 lte(\_,In) & yes & yes & yes & subset(In,In) & yes & yes & yes\\
\cline{1-8}
map(In,\_) & yes & yes & yes &subset (In,Out) & yes & no &\\
\cline{1-8}
map(\_,In) & yes & yes & yes & subset (Out,In) & yes & no &\\
\cline{1-8}
member(\_,In) & yes & yes & yes &sum(\_,In,\_) & yes & yes & yes\\
\cline{1-8}
member(In,Out) & yes & no &  &sum(\_,\_,In) & yes & yes & yes \\
\cline{1-8}
mergesort(In,\_) & yes & yes & no & sum(In,Out,Out) & yes & no & \\
\cline{1-8}
mergesort(Out,In) & yes & no &  &type(In,In,Out) & no & yes & no \\
\cline{1-8}
mergesort\_variant(\_,\_,In) & yes & yes & yes &type(In,Out,Out) & no & no &\\
\hline\hline
\end{tabular}
\end{table}

\begin{table}
\caption{Programs from  DPPD's Collection}
\label{2}
\centering
\begin{tabular}{lccc||lccc}
\hline\hline
& {\bf NM} & {\bf IT} &   {\bf  QR} & &  {\bf NM} & {\bf IT} &   {\bf  QR}\\
\cline{1-8}
applast(In,In,Out) & yes  & yes & yes & match\_app(In,Out) & yes & no &\\
\cline{1-8}
applast(Out,\_,\_) & yes  & no &  &max\_lenth(In,Out,Out) & yes & yes & yes\\
\cline{1-8}
applast(\_,Out,\_) & yes  & no &  &memo\_solve(In,Out) & yes & yes & no\\
\cline{1-8}
contains(\_,In) & yes & yes & yes & power(In,In,In,Out) & yes & yes & yes\\
\cline{1-8}
contains(In,Out) & yes & no &  &  prune(In,\_) & yes & yes & yes\\
\cline{1-8}
depth(In,In) & yes & yes & yes & prune(\_,In) & yes & yes & yes\\
\cline{1-8}
depth(In,Out) & yes & yes & no & relative (In,\_) & yes & no & \\
\cline{1-8}
depth(Out,In) & yes & no &  &  relative(\_,In) & yes & no & \\
\cline{1-8}
duplicate(In,Out) & yes  & yes & yes  &rev\_acc(In,In,Out) & yes & yes & yes \\
\cline{1-8}
duplicate(Out,In) & yes  & yes & yes  & rotate(In,\_)  & yes & yes & yes\\
\cline{1-8}
flipflip(In,Out) & yes & yes & yes  & rotate(\_,In)  & yes & yes & yes\\
\cline{1-8}
flipflip(Out,In) & yes & yes & yes &solve(\_,\_,\_)  & yes & no & \\
\cline{1-8}
generate(In,In,Out)  & yes & no &  &ssupply(In,In,Out) & yes & yes & yes\\
\cline{1-8}
liftsolve(In,Out) & yes & no &  & trace(In,In,Out)   & yes & yes & yes\\
\cline{1-8}
liftsolve(Out,In) & yes & no &  & transpose(\_,In) & yes & yes & yes\\
\cline{1-8}
liftsolve(In,In) & yes & yes & yes  & transpose(In,Out) & yes & no &\\
\cline{1-8}
 match\_app(\_,In) & yes & yes & yes& unify(In,In,Out) & yes & no & \\
\hline\hline
\end{tabular}
\end{table}

\begin{table}
\caption{Programs from  Lindenstrauss's Collection}
\label{3}
\centering
\begin{tabular}{lccc||lccc}
\hline\hline
& {\bf NM} & {\bf IT} &   {\bf  QR} & &  {\bf NM} & {\bf IT} &   {\bf  QR}\\
\cline{1-8}
ack(In,In,\_) & yes & yes & no & least(In,\_) &  yes & yes & yes\\
\cline{1-8}
concatenate(In,\_,\_) & yes & yes & yes &  least(\_,In) &  yes & yes & yes\\
\cline{1-8}
concatenate(\_,\_,In) & yes & yes & yes & normal\_form(In,\_) & yes & no & \\
\cline{1-8}
concatenate(\_,In,\_) & yes & no &  &normal\_form(\_,In) & yes & no &\\
\cline{1-8}
descendant(In,\_) & yes & no &  & queens(\_,Out) & yes & yes & no\\
\cline{1-8}
descendant(\_,In) & yes & no &  &queens(\_,In) & yes & yes & yes\\
\cline{1-8}
deep(In,\_) & yes & yes & yes &poss(In)  &  yes & yes & yes\\
\cline{1-8}
deep(Out,\_)  & yes & no & &poss(Out) &  yes & no &\\
\cline{1-8}
credit(In,\_) & yes & yes & yes &rewrite(In,\_) & yes & yes & yes\\
\cline{1-8}
credit(\_,In) & yes & yes & yes & rewrite(\_,In) & yes & yes & yes\\
\cline{1-8}
holds(\_,Out) & yes & no &  & transform(\_,\_,\_,Out) & yes & no & \\
\cline{1-8}
holds(\_,In) & yes & yes & yes & transform(\_,\_,\_, In) & yes & yes & yes\\
\cline{1-8}
huffman(In,\_) &  yes & yes & no &twoleast(In,\_) & yes & yes & yes\\
\cline{1-8}
huffman(\_,In) &  yes & no &  &twoleast(\_,In) & yes & yes & yes\\
\hline\hline
\end{tabular}
\end{table}

\begin{table}
\caption{Programs from  Hermenegildo's Collection}
\label{4}
\centering
\begin{tabular}{llccc}
\hline\hline
& & {\bf NM} & {\bf IT} &   {\bf  QR} \\
\cline{1-5}
aiakl.pl  & init\_vars(In,In,Out,Out) & yes & yes & yes \\
\cline{1-5}
ann.pl & analyze\_all(In,Out) & yes  & yes & yes\\
\cline{1-5}
bid.pl &
bid(In,Out,Out,Out) & yes  & yes & yes\\
\cline{1-5}
 boyer.pl &
tautology(In) & yes & no & \\
\cline{1-5}
browse.pl &
investigate(In,Out) & yes & yes & yes\\
\cline{1-5}
fib.pl &
fib(In,Out) & yes & no & \\
\cline{1-5}
fib\_add.pl &
fib(In,Out) & yes & yes & yes\\
\cline{1-5}
hanoiapp.pl & 
shanoi(In,In,In,In,Out) & yes & no & \\
\cline{1-5}
hanoiapp\_suc.pl &
shanoi(In,In,In,In,Out) & yes & yes & yes \\
\cline{1-5}
mmatrix.pl & 
mmultiply(In,In,Out) & yes & yes & yes\\
\cline{1-5}
occur.pl &
occurall(In,In,Out) & yes & yes & yes \\
\cline{1-5}
peephole.pl & 
peephole\_opt(In,Out) & yes  & yes & yes \\
\cline{1-5}
progeom.pl & 
pds(In,Out) & yes  & yes & yes \\
\cline{1-5}
rdtok.pl &
read\_tokens(In,Out) & yes & no & \\
\cline{1-5}
read.pl & 
parse(In,Out) & yes & no & \\
\cline{1-5}
serialize.pl &
serialize(In,Out) & yes & yes & no\\
\cline{1-5}
tak.pl &
tak(In,In,In,Out) & yes &  no & \\
\cline{1-5}
tictactoe.pl & 
play(In) & yes & no & \\
\cline{1-5}
warplan.pl &
plans(In,In) & yes & no & 
\\
\hline\hline
\end{tabular}
\end{table}

\begin{table}
\caption{Input terminatining but non-quasi recurrent Programs}
\label{5}
\centering
\begin{tabular}{lcc}
\hline\hline
 & {\bf SM} & {\bf IR}\\
\cline{1-3}
mergesort(In,\_) & yes & no \\
\cline{1-3}
 qsort(In,\_) & yes &  no\\
\cline{1-3}
 type(In,In,Out) & no  & no \\
\cline{1-3}
depth(In,Out) & yes & no \\
\cline{1-3}
memo\_solve(In,Out) & no & no\\
\cline{1-3}
 ack(In,In,\_) & yes & no\\
\cline{1-3}
huffman(In,\_) &  no & no\\
\cline{1-3}
 queens(\_,Out) & no & no\\
\cline{1-3}
serialize(In,Out) & no & no\\
\hline\hline
\end{tabular}
\end{table}

\end{document}